# Ward-Takahashi Identities, Magnetic Anomalies, and the Anticommutation Properties of the Fermion-Boson Vertex

## Jay R. Yablon[*]

910 Northumberland Drive

Schenectady, New York, 12309-2814

**Abstract:**

**It is well-known that following summing Feynman graphs, the fermion-boson coupling vertex is modified according to $\gamma^\mu \Rightarrow \Gamma^\mu = \gamma^\mu + \Lambda^\mu$, with $\Lambda^\mu$ representing non-divergent perturbative corrections. Here, we calculate the anticommutators specified by $\Gamma^{\mu\nu} \equiv \frac{1}{2}\{\Gamma^\mu, \Gamma^\nu\}$, and then explore some consequences of employing these as a metric tensor $g^{\mu\nu} \equiv \Gamma^{\mu\nu}(p',q,p)$ in momentum space. The challenge is that $\Gamma^\mu$ and $\Gamma^{\mu\nu}$ must then be introduced in place of $\gamma^\mu$ and $\eta^{\mu\nu}$ throughout the Lagrangian density, denoted $\mathcal{L}$', resulting in what appears, superficially, to be different physics from what is known and observed. However, with a suitable reparameterization of fermion rest masses $m'$, interaction charges $e'$ and momentum vectors $p^{\mu'}$ into their observed counterparts $m$, $e$ and $p^\mu$, it turns out that $\mathcal{L}$' can be made to describe physics identical to that of the customary QED Dirac Lagrangian density $\mathcal{L}$ at low photon momentum $q^\mu \to 0$, including the observed magnetic anomaly. That is, we prove that one is able to obtain $\mathcal{L}(m,e,p^\mu)$ = $\mathcal{L}'(m',e',p^{\mu'})$ for $q^\mu \to 0$. We find through the Ward-Takahashi identity, as summarized in Figure 2, that interaction vertexes are proportional to the difference between the ordinary and covariant momentum-space derivatives of the metric tensor, and thus an indicator of curvature. Finally, we obtain additional non-longitudinal terms in the Ward-Takahashi relation.**

---

[*] jyablon@nycap.rr.com

**Contents**

# 1. Introduction

The successful explanation of the electron's intrinsic magnetic moment, and then of its anomalous magnetic moment to eight- or nine-digit accuracy, establish quantum field theory (QFT) as the most accurate description of nature attained to date. First, the Dirac equation predicts a gyromagnetic ratio $g = 2$ for the intrinsic magnetic moment $\mu = g\frac{e\hbar}{m}\frac{\mathbf{S}}{2} = 2\frac{e\hbar}{m}\frac{\mathbf{S}}{2}$ of the electron, as a very good zero-order approximation. [1] Second, Schwinger's 1948 discovery [2] shows the gyromagnetic ratio in first order to be given by $\frac{g}{2} = 1 + \frac{a}{2\pi} + \ldots = 1.00116141$, where $a = 1/137.03599911$ [3] is the modern-day low-probe-energy value of the running electromagnetic (fine structure) coupling. Since Schwinger, even more-precise fits with phenomenological data have been obtained using higher-order electromagnetic loop diagrams as well as those for weak and strong interactions.

This understanding of anomalous magnetic moments originates from summing Feynman graphs for whatever interactions are being considered to whatever order is being considered, capturing these results in the form factors $F_1, F_2$, (see [4], equation (11.3.29)), and then modifying the Dirac matrices $\gamma^\mu$ according to $\gamma^\mu \Rightarrow \Gamma^\mu(p,q,p') = \gamma^\mu + \Lambda^\mu(p,q,p')$, for the fermion-boson vertex. Above, $\Lambda^\mu$ represents a finite (non-divergent) correction to this vertex, see, e.g., [5], pp. 343-345, and $p$, $p'$ and $q$ represent incoming and outgoing fermion momentum, and photon momentum, respectively. The Lagrangian density used to obtain the anomalous magnetic moment is given by $\mathcal{L} = i\bar\psi\gamma^\mu\partial_\mu\psi + e\bar\psi\Gamma^\mu A_\mu\psi - m\bar\psi\psi - \frac{1}{4}F^{\mu\nu}F_{\mu\nu}$, which corresponds to Dirac's momentum space equation $(\not{p}-m)u(p) = e\Gamma^\sigma A_\sigma u(p)$ and Maxwell's equation $J^\mu = F^{\sigma\mu}{}_{;\sigma}$. It is important to observe that the correction $\gamma^\mu \Rightarrow \Gamma^\mu$ is applied in the fermion-photon interaction term $e\bar\psi\Gamma^\mu A_\mu\psi$, but *not in the kinetic term* $i\bar\psi\gamma^\mu\partial_\mu\psi$.

Because the Dirac $\gamma^\mu$ are related to the contravariant Minkowski metric tensor $\eta^{\mu\nu}$ via the anticommutators $\eta^{\mu\nu} \equiv \frac{1}{2}(\gamma^\mu\gamma^\nu + \gamma^\nu\gamma^\mu) = \frac{1}{2}\{\gamma^\mu,\gamma^\nu\}$, it is of interest to at least inquire, *mathematically*, about the anticommutation properties of the corrected matrices $\Gamma^\mu$, by defining and then calculating the analogous anticommutators $\Gamma^{\mu\nu} \equiv \frac{1}{2}\{\Gamma^\mu,\Gamma^\nu\}$. What is especially interesting, is to then consider what the *physics* would be, *if, hypothetically for now,* this $\Gamma^{\mu\nu}$, which approaches $\eta^{\mu\nu}$ in the $\Gamma^\mu \to \gamma^\mu$ limit, were to be employed as the contravariant metric tensor $g^{\mu\nu} \equiv \Gamma^{\mu\nu}$ of a gravitational field. If such a connection can be established, it may become possible to draw closer together, the seemingly-disparate theories of quantum fields for which the Dirac-Schwinger magnetic moment explanation is a central phenomenological foundation, and geometrically-based gravitation. That is the subject of this paper.

# 2. Calculation of the Anticommutator Matrices $\Gamma^{\mu\nu} \equiv \frac{1}{2}\{\Gamma^\mu,\Gamma^\nu\}$

We begin with equation (11.3.29) from Weinberg's definitive treatise *The Quantum Theory of Fields* [4], which is reproduced in (2.1) below: ($\sigma^{\mu\nu} \equiv \frac{1}{2}i[\gamma^\mu\gamma^\nu - \gamma^\nu\gamma^\mu]$)

$$\bar{u}(\mathbf{p}',\sigma')\Gamma^{\mu}(p',p)u(\mathbf{p},\sigma) = \bar{u}(\mathbf{p}',\sigma')\{\gamma^{\mu}F_1(q^2) + \tfrac{1}{2}i\sigma^{\mu\nu}(p'-p)_\nu F_2(q^2)\}u(\mathbf{p},\sigma). \tag{2.1}$$

The form factors $F$ are related by: (see [4], equations (10.3.30), (10.6.17), (10.6.18) respectively)

$$F_1(q^2) = F(q^2) + G(q^2) \tag{2.2}$$

$$F(q^2) = F_1(q^2) + 2mF_2(q^2). \tag{2.3}$$

$$G(q^2) = -2mF_2(q^2). \tag{2.4}$$

Schwinger's magnetic moment calculation, which is a representative example of the use of these form factors, and which applies for $\alpha \ll 1$ when only electromagnetic interactions are considered, employs the particular form factors:

$$F_1 = 1; \quad F_2 = \frac{\alpha}{2\pi}\frac{1}{m}; \quad G = -\frac{\alpha}{\pi}; \quad F = 1+\frac{\alpha}{\pi}; \quad F_1 + F_2 m = 1+\frac{\alpha}{2\pi}. \tag{2.5}$$

It is clear from this that $F_1 + mF_2 = g/2$ specifies the gyromagnetic ratio. (Contrast equations (9.136), (9.138) in [5].) Now, we turn to the anticommutator calculation.

From (2.1), we define $\Lambda^\mu$, and extract $\Gamma^\mu$, as such:

$$\Gamma^\mu \equiv \gamma^\mu + \Lambda^\mu = F_1\gamma^\mu + \tfrac{1}{2}F_2 i\sigma^{\mu\nu}(p'-p)_\nu. \tag{2.6}$$

We then *define* the anticommutator $\Gamma^{\mu\nu}$ as the symmetric mathematical object:

$$\Gamma^{\mu\nu} \equiv \tfrac{1}{2}\{\Gamma^\mu\Gamma^\nu + \Gamma^\nu\Gamma^\mu\} = \tfrac{1}{2}\{\Gamma^\mu,\Gamma^\nu\}. \tag{2.7}$$

First, substituting (2.6) into (2.7), and using $2\eta^{\mu\nu} = \gamma^\mu\gamma^\nu + \gamma^\nu\gamma^\mu$, we get:

$$\Gamma^{\mu\nu} = F_1^2 \eta^{\mu\nu} - \tfrac{1}{8}F_1 F_2 \begin{Bmatrix} +\gamma^\mu\gamma^\nu\gamma^\sigma - \gamma^\mu\gamma^\sigma\gamma^\nu + \gamma^\nu\gamma^\sigma\gamma^\mu - \gamma^\sigma\gamma^\nu\gamma^\mu \\ +\gamma^\mu\gamma^\sigma\gamma^\nu - \gamma^\sigma\gamma^\mu\gamma^\nu + \gamma^\nu\gamma^\mu\gamma^\sigma - \gamma^\nu\gamma^\sigma\gamma^\mu \end{Bmatrix}(p'-p)_\sigma$$
$$+ \tfrac{1}{32}F_2^2 \begin{Bmatrix} +\gamma^\mu\gamma^\sigma\gamma^\nu\gamma^\tau - \gamma^\sigma\gamma^\mu\gamma^\nu\gamma^\tau - \gamma^\mu\gamma^\sigma\gamma^\tau\gamma^\nu + \gamma^\sigma\gamma^\mu\gamma^\tau\gamma^\nu \\ +\gamma^\nu\gamma^\tau\gamma^\mu\gamma^\sigma - \gamma^\tau\gamma^\nu\gamma^\mu\gamma^\sigma - \gamma^\nu\gamma^\tau\gamma^\sigma\gamma^\mu + \gamma^\tau\gamma^\nu\gamma^\sigma\gamma^\mu \end{Bmatrix}(p'-p)_\sigma(p'-p)_\tau \tag{2.8}$$

We can factor out all of the $(p'-p)_\sigma$ in this way because the $\gamma^\sigma$ with which these are summed hold the commutation position for $p'-p\!\!\!/ = \gamma^\sigma(p'-p)_\sigma$.

For the first-order term with coefficient $-\tfrac{1}{8}F_1 F_2$, we move all of the $\gamma^\sigma$ all the way to the right by repeatedly applying $\gamma^\sigma\gamma^\mu = 2\eta^{\mu\sigma} - \gamma^\mu\gamma^\sigma$. We find that all terms cancel identically:

$$+\gamma^\mu\gamma^\nu\gamma^\sigma - \gamma^\mu\gamma^\sigma\gamma^\nu + \gamma^\nu\gamma^\sigma\gamma^\mu - \gamma^\sigma\gamma^\nu\gamma^\mu + \gamma^\mu\gamma^\sigma\gamma^\nu - \gamma^\sigma\gamma^\mu\gamma^\nu + \gamma^\nu\gamma^\mu\gamma^\sigma - \gamma^\nu\gamma^\sigma\gamma^\mu = 0. \tag{2.9}$$

For the second-order term with coefficient $\frac{1}{32}F_2^2$, we repeatedly apply $\gamma^\sigma \gamma^\mu = 2\eta^{\mu\sigma} - \gamma^\mu \gamma^\sigma$ with suitable indexes, to move $\gamma^\mu$ and $\gamma^\nu$ into the middle, sandwiched by $\gamma^\sigma$ and $\gamma^\tau$, in the form of the $\gamma^\sigma \gamma^\mu \gamma^\nu \gamma^\tau$ term. After several iterations, this all reduces to:

$$-8\gamma^\sigma \eta^{\mu\nu} \gamma^\tau + 8\eta^{\sigma\mu}\eta^{\nu\tau}$$
$$= +\gamma^\mu \gamma^\sigma \gamma^\nu \gamma^\tau - \gamma^\sigma \gamma^\mu \gamma^\nu \gamma^\tau - \gamma^\mu \gamma^\sigma \gamma^\tau \gamma^\nu + \gamma^\sigma \gamma^\mu \gamma^\tau \gamma^\nu \quad (2.10)$$
$$+ \gamma^\nu \gamma^\tau \gamma^\mu \gamma^\sigma - \gamma^\tau \gamma^\nu \gamma^\mu \gamma^\sigma - \gamma^\nu \gamma^\tau \gamma^\sigma \gamma^\mu + \gamma^\tau \gamma^\nu \gamma^\sigma \gamma^\mu$$

Returning to (2.8) using (2.9) and (2.10), this means that the covariant anticommutator is:

$$\Gamma_{\mu\nu} = F_1^2 \eta_{\mu\nu} - \tfrac{1}{4}F_2^2 \left( \eta_{\mu\nu}(p'-p)(p'-p) - \eta_{\sigma\mu}\eta_{\nu\tau}(p'-p)^\sigma (p'-p)^\tau \right). \quad (2.11)$$

This matrix, $\Gamma_{\mu\nu}$, is the end result of this anticommutator calculation. For now, $\Gamma_{\mu\nu}$ are merely mathematical objects, namely, the anticommutators $\Gamma^{\mu\nu} \equiv \tfrac{1}{2}\{\Gamma^\mu \Gamma^\nu + \Gamma^\nu \Gamma^\mu\}$ of the $\Gamma^\mu \equiv \gamma^\mu + \Lambda^\mu$ applied in the fermion-photon interaction term $e\overline{\psi}\Gamma^\mu A_\mu \psi$ of the Lagrangian density $\mathcal{L}$.

If we want to think about this in a specific, simple case, we might apply the Schwinger form factors in (2.5) to write:

$$\Gamma_{\mu\nu} = \eta_{\mu\nu} - \frac{\alpha^2}{16\pi^2 m^2}\left[\eta_{\mu\nu}(p'-p)(p'-p) - \eta_{\sigma\mu}\eta_{\nu\tau}(p'-p)^\sigma (p'-p)^\tau \right]. \quad (2.12)$$

It will also be helpful in our discussion to refer to: (from Weinberg [4], equation (11.3.31))

$$F_1(q^2) \cong 1 + \frac{e^2}{24\pi^2}\left(\frac{q^2}{m^2}\right)\left[\ln\left(\frac{\mu^2}{m^2}\right) + \frac{2}{5} + \frac{3}{4}\right]. \quad (2.13)$$

Now, looking at (2.11) and (2.12), we see that where $F_1 \cong 1$ and $F_2 \cong 0$, $\Gamma_{\mu\nu}$ approximates the Minkowski metric $\eta_{\mu\nu}$ tensor of flat spacetime very closely, $\Gamma_{\mu\nu} \cong \eta_{\mu\nu}$. One is thus motivated to at least ask: what would it mean *if, hypothetically,* we tried to associate $\Gamma_{\mu\nu}$ with the metric tensor for a non-zero gravitational field by setting $\Gamma_{\mu\nu} = g_{\mu\nu}$, where $g_{\mu\nu} \neq \eta_{\mu\nu}$ contains a gravitational field? What would the physics look like? Would this contradict any important pedagogical or experimental results? Can we retrieve from this, the known magnetic anomaly? Does this explain anything new?

## 3. The Ward-Takahashi Identity and Covariant Differentiation in Momentum Space

*If, as a hypothesis for exploration,* we were to interpret $\Gamma_{\mu\nu}$ as a metric tensor $\Gamma_{\mu\nu} = g_{\mu\nu}$, then we *must* use this tensor to raise and lower indexes and define matrix inverses, i.e., we must require that $\Gamma^{\mu\sigma}\Gamma_{\nu\sigma} = \delta^\mu{}_\nu$. Further, $\Gamma_{\mu\nu}$ must be involved in all couplings between vectors and tensors in spacetime, and $\Gamma^\mu$ must appear in place of $\gamma^\mu$ at *all* vertexes, not only the fermion-boson vertex $e\bar\psi\Gamma^\mu A_\mu\psi$. Additionally, a $g_{\mu\nu}(p,q,p') \equiv \Gamma_{\mu\nu}$ would *not* be the usual function of spacetime coordinates, $g_{\mu\nu}(x^\sigma)$, but rather of momentum vectors $p^\mu, q^\mu, p^{\mu'}$. Thus, a $g_{\mu\nu}(p,q,p')$ as defined in (2.11) would be a *metric tensor in momentum space*. We shall explore this latter feature momentarily. Finally, because the $g_{\mu\nu}(p,q,p')$ contain the Dirac $\gamma^\mu$ within, these can be used to operate on Dirac spinors, in the general form $g_{\mu\nu}\psi$, and one must pay careful attention to their commutivity properties. This will be explored further in section 8.

Now, $\Gamma_{\mu\nu}$ in (2.11) is not the only form of these anticommutators. Starting with the Dirac equation $(\gamma^\mu p_\mu - m)u(p) = 0$ and its adjoint $\bar u(p')(\gamma^\mu p'_\mu - m) = 0$ for free fermions, it is a common technique to combine $\eta^{\mu\nu} = \tfrac{1}{2}[\gamma^\mu\gamma^\nu + \gamma^\nu\gamma^\mu]$ and $\sigma^{\mu\nu} = \tfrac{1}{2}i[\gamma^\mu\gamma^\nu - \gamma^\nu\gamma^\mu]$ into the expression $\gamma^\mu\gamma^\nu = -\gamma^\mu\gamma^\nu + 2\eta^{\mu\nu} - 2i\sigma^{\mu\nu}$, and thereby to rewrite Dirac's equation in the form $\frac{1}{m}(p^\nu + i\sigma^{\mu\nu}p_\mu)u(p) = \gamma^\nu u(p)$ and the adjoint as $\frac{1}{m}\bar u(p')(p'^\nu - i\sigma^{\mu\nu}p'_\mu) = \bar u(p')\gamma^\nu$. These are then combined to form $\bar u(p')\gamma^\mu u(p) = \frac{1}{2m}\bar u(p')\left((p'+p)^\mu + i\sigma^{\mu\nu}(p'-p)_\nu\right)u(p)$, see [5], equation (9.136). Stripping off the spinors and rearranging this as $\tfrac{1}{2}i\sigma^{\mu\nu}(p'-p)_\nu = m\gamma^\mu - \tfrac{1}{2}(p'+p)^\mu$, we may use this "Gordon decomposition" to rewrite the vertex function (2.6) as:

$$\Gamma^\mu = (F_1 + F_2 m)\gamma^\mu - \tfrac{1}{2}F_2(p'+p)^\mu. \tag{3.1}$$

Again using $\Gamma^{\mu\nu} \equiv \tfrac{1}{2}\{\Gamma^\mu\Gamma^\nu + \Gamma^\nu\Gamma^\mu\}$ in (2.7), the calculation is much more straightforward than the one in section 2. One arrives easily at the alternative, contravariant anticommutator:

$$\Gamma^{\mu\nu} = (F_1 + F_2 m)^2\eta^{\mu\nu} - \tfrac{1}{2}(F_1 + F_2 m)F_2\left(\gamma^\mu(p'+p)^\nu + \gamma^\nu(p'+p)^\mu\right) + \tfrac{1}{4}F_2^2(p'+p)^\mu(p'+p)^\nu. \tag{3.2}$$

Equations (2.11) and (3.2) are totally equivalent via the Gordon decomposition, but for the fact that (2.11) is written in covariant and (3.2) in contravariant form. This is done to place the momentum terms $p'$, $p$ into contravariant form $p'^\mu$, $p^\mu$ so their components are most easily specified. Setting $q^\mu \equiv (p'-p)^\mu$ where $q^\mu$ is a massless vector boson (e.g. photon) momentum, so that $(p'+p)^\mu = 2p^\mu + q^\mu$, (3.1) and (3.2) become, respectively:

$$\Gamma^\mu = (F_1 + F_2 m)\gamma^\mu - \tfrac{1}{2}F_2(2p^\mu + q^\mu), \tag{3.3}$$

$$\Gamma^{\mu\nu} = (F_1 + F_2 m)^2\eta^{\mu\nu} - \tfrac{1}{2}(F_1 + F_2 m)F_2\left(\gamma^\mu(2p^\nu + q^\nu) + \gamma^\nu(2p^\mu + q^\mu)\right) \\ + \tfrac{1}{4}F_2^2(4p^\mu p^\nu + 2p^\mu q^\nu + 2p^\nu q^\mu + q^\mu q^\nu) \tag{3.4}$$

It should be noted that (3.2), (3.4), with terms like $(p'+p)^\mu = 2p^\mu + q^\mu$, lend themselves naturally to representing the momentum $p^\mu$ of a fermion interacting with a vector boson with momentum $q^\mu$ or to representing just $p^\mu$ alone, when $q^\mu \to 0$. By contrast, (2.11), with terms like $\not{q} = \not{p}' - \not{p}$ and $q^\sigma = (p'-p)^\sigma$, lends most naturally to representing a vector boson momentum $q^\mu$, independently of any interaction with a fermion.

It is now extremely instructive to differentiate (3.4) with respect to the incoming fermion momentum $p^\nu$. But, we need to be careful because of the appearance of terms $p^\mu$ in addition to $p^\nu$. For $p^\nu$, it is clear that $\partial p^\nu / \partial p^\nu = 1$, as is seen for example, by recognizing that the second-order Ward identity $\Gamma^\mu(p,0,p) = \dfrac{\partial S'^{-1}_F}{\partial p_\mu} = \dfrac{\partial(\gamma^\mu p_\mu - m - \Sigma)}{\partial p_\mu} = \gamma^\mu - \dfrac{\partial \Sigma}{\partial p_\mu}$ includes $\partial p_\mu / \partial p_\mu = 1$, where $\Sigma$ is the self-energy bubble in the fermion line $\overset{p}{\rule{1em}{0.4pt}}\!\bigcirc\!\overset{p}{\rule{1em}{0.4pt}}$ (see [5], equations (7.112), (7.113), (7.121), (7.122), and section 7.4 generally). But what of terms such as $\partial p^\mu / \partial p^\nu$? Here, we apply the tensor relationship $\partial p^\mu / \partial p^\nu = \delta^\mu{}_\nu$, which is to say that $\partial p^0 / \partial p^0 = 1$, $\partial p^1 / \partial p^1 = 1$, etc., but $\partial p^0 / \partial p^1 = 0$, etc., due to each of the four components of $p^\mu$ being independent. Where we must exercise special care is never to contract $\partial p^\mu / \partial p^\nu = \delta^\mu{}_\nu$, because this would yield $\partial p^\mu / \partial p^\mu = \delta^\mu{}_\mu = 4$, which is incorrect, because $\partial p^\mu / \partial p^\mu = 1$, as just discussed.

With this in mind, we differentiate (3.4) with respect to $p^\nu$, showing the full calculation:

$$\dfrac{\partial \Gamma^{\mu\nu}(p,q,p')}{\partial p^\nu}$$
$$= -(F_1 + F_2 m)F_2\left(\gamma^\mu \dfrac{\partial p^\nu}{\partial p^\nu} + \gamma^\nu \dfrac{\partial p^\mu}{\partial p^\nu}\right) + \tfrac{1}{4}F_2{}^2\left(4\dfrac{\partial p^\nu}{\partial p^\nu}p^\mu + 4\dfrac{\partial p^\mu}{\partial p^\nu}p^\nu + 2\dfrac{\partial p^\nu}{\partial p^\nu}q^\mu + 2\dfrac{\partial p^\mu}{\partial p^\nu}q^\nu\right), \quad (3.5)$$
$$= -(F_1 + F_2 m)F_2(\gamma^\mu + \gamma^\nu \delta^\mu{}_\nu) + \tfrac{1}{4}F_2{}^2(4p^\mu + 4\delta^\mu{}_\nu p^\nu + 2q^\mu + 2\delta^\mu{}_\nu q^\nu)$$
$$= -2F_2\left[(F_1 + F_2 m)\gamma^\mu - \tfrac{1}{2}F_2(2p^\mu + q^\mu)\right] = -2F_2\Gamma^\mu(p,q,p')$$

where we have made use of (3.3) in the final step. One may derive the similar expressions $\partial \Gamma^{\mu\nu}(p,q,p')/\partial p'^\nu = -2F_2\Gamma^\mu(p,q,p')$; $\partial \Gamma^{\mu\nu}(p,q,p')/\partial q^\nu = -F_2\Gamma^\mu(p,q,p')$.

*If* we now regard $\Gamma_{\mu\nu}$ as a metric tensor $\Gamma_{\mu\nu} = g_{\mu\nu}$ in momentum space, then the relation $\partial g^{\mu\nu}(p,q,p')/\partial p^\nu = -2F_2\Gamma^\mu(p,q,p')$ found in (3.5) makes it possible to write the Ward-Takahashi identity *in terms of this metric tensor* as:

$$-\dfrac{1}{2}\dfrac{q_\mu}{F_2}\dfrac{\partial g^{\mu\nu}(p,q,p')}{\partial p^\nu} = q_\mu \Gamma^\mu(p,q,p') = S'^{-1}_F(p') - S'^{-1}_F(p), \quad (3.6)$$

and to write the $q_\mu \to 0$ Ward identity, which applies at $(p,0,p)$, as (see [5], equations (7.111), (7.112), and section 7.4 generally):

$$-\frac{1}{2}\frac{1}{F_2}\frac{\partial g^{\mu\nu}(p,0,p)}{\partial p^\nu} = \Gamma^\mu(p,0,p) = \frac{\partial S'^{-1}_F}{\partial p_\mu}. \tag{3.7}$$

We are thereby able to relate these identities directly to the *ordinary* derivative of the metric tensor $g^{\mu\nu}(p,q,p')$ in momentum space. We shall momentarily consider *covariant* derivatives.

First, let's simplify notation. In spacetime, "," and ";" denote ordinary and covariant derivatives, for example, $g^{\mu\nu}{}_{,\alpha} \equiv \partial g^{\mu\nu}/\partial x^\alpha$ and $g^{\mu\nu}{}_{;\alpha} \equiv g^{\mu\nu}{}_{,\alpha} + \Gamma^\mu{}_{\sigma\alpha}g^{\sigma\nu} + \Gamma^\nu{}_{\sigma\alpha}g^{\mu\sigma}(\equiv 0)$. In momentum space, we similarly introduce "/" and "⟩" to denote "ordinary" and "covariant" differentiation *with respect to incoming fermion momentum $p^\mu$*, thus rewriting (3.6) and (3.7) with a "/" as:

$$q_\mu g^{\mu\nu}(p,q,p')_{/\nu} = -2F_2 q_\mu \Gamma^\mu(p,q,p') = -2F_2\left(S'^{-1}_F(p') - S'^{-1}_F(p)\right), \tag{3.8}$$

$$g^{\mu\nu}(p,0,p)_{/\nu} = -2F_2\Gamma^\mu(p,0,p) = -2F_2 S'^{-1/\mu}_F. \tag{3.9}$$

The Ward-Takahashi identities, of course, are vital to order-by-order renormalizability in gauge theory, so, it is desirable to make these as fundamental as possible to how we understand $g^{\mu\nu}$ in momentum space. In fact, just as the *covariant* derivative $g^{\mu\nu}{}_{;\alpha} \equiv 0$ in spacetime, we can use (3.5), compactly written as $g^{\mu\nu}{}_{/\nu} + 2F_2\Gamma^\mu = 0$, and which applies in general for $(p,q,p')$, to *define* the contracted covariant derivative in momentum space to be $g^{\mu\nu}{}_{\rangle\nu} \equiv 0$. Thus, we *define*:

$$g^{\mu\nu}(p,q,p')_{\rangle\nu} \equiv 0 = g^{\mu\nu}(p,q,p')_{/\nu} + 2F_2\Gamma^\mu(p,q,p'). \tag{3.10}$$

Then, we rewrite (3.8) and (3.9) in terms of this contracted *covariant* derivative, as:

$$\begin{aligned}q_\mu g^{\mu\nu}(p,q,p')_{\rangle\nu} &\equiv 0 = q_\mu g^{\mu\nu}(p,q,p')_{/\nu} + 2F_2 q_\mu \Gamma^\mu(p,q,p') \\ &= -2F_2 q_\mu \Gamma^\mu(p,q,p') + 2F_2\left(S'^{-1}_F(p') - S'^{-1}_F(p)\right)\end{aligned} \tag{3.11}$$

$$g^{\mu\nu}(p,0,p)_{\rangle\nu} \equiv 0 = g^{\mu\nu}(p,0,p)_{/\nu} + 2F_2\Gamma^\mu(p,0,p) = -2F_2\Gamma^\mu(p,0,p) + 2F_2 S'^{-1/\mu}_F. \tag{3.12}$$

*This embeds the Ward-Takahashi identities into the momentum space metric tensor directly through how we define covariant differentiation.* Of course, in spacetime, the Christoffel connections $\Gamma^\mu{}_{\alpha\beta} \equiv \frac{1}{2}g^{\mu\nu}\left(g_{\nu\alpha,\beta} + g_{\beta\nu,\alpha} - g_{\alpha\beta,\nu}\right)$ are used to define covariant derivatives, which, for a second rank tensor such as $g^{\mu\nu}$, are given by $g^{\mu\nu}{}_{;\beta} = g^{\mu\nu}{}_{,\beta} + \Gamma^\mu{}_{\alpha\beta}g^{\alpha\nu} + \Gamma^\nu{}_{\alpha\beta}g^{\mu\alpha}(\equiv 0)$. Thus, we must relate (3.10) to a more-general definition of covariant differentiation in momentum space. More to the point, we must *define* covariant differentiation in momentum space, via a suitable set of connections $\Gamma^\mu{}_{\alpha\beta}$, *such that (3.10) is satisfied identically*. If we can do so, then (3.11) would tell us something very fundamental about interaction vertexes in momentum space:

Specifically, we now draw (3.11) in terms of its associated Feynman graphs, as in Figure 1 below (see [5], Figure 7.4):

$$q_\mu \left[ g^{\mu\nu}(p,q,p')_{;\nu} - g^{\mu\nu}(p,q,p')_{/\nu} \right] \equiv 2F_2 q_\mu \begin{array}{c} q \\ \diagup \diagdown \\ p \quad p+q \end{array} \equiv 2F_2 \left[ \overline{p+q} \underset{\circ}{\phantom{p+q}} \overline{p+q}^{-1} - \overline{p} \underset{\circ}{\phantom{p}} \overline{p}^{-1} \right]$$

Figure 1

Then, we simply divide out the $q_\mu$, to obtain:

$$g^{\mu\nu}(p,q,p')_{;\nu} - g^{\mu\nu}(p,q,p')_{/\nu} \equiv 2F_2 \begin{array}{c} q \\ \diagup \diagdown \\ p \quad p+q \end{array}$$

Figure 2

Figure 2 now reveals that *the difference between the contracted covariant and ordinary derivatives of the momentum space metric tensor, is equal to the scattering vertex times twice the form factor $F_2$*. We know, however, that the Riemann curvature tensor $R^\alpha{}_{\beta\mu\nu}$ in spacetime is *defined* as a measure of the degree to which covariant derivatives do not commute, that is, by $R^\alpha{}_{\beta\mu\nu} A_\alpha \equiv A_{\beta;\mu;\nu} - A_{\beta;\nu;\mu}$. Where the covariant and ordinary derivatives are identical, for example, $g^{\mu\nu}{}_{;\nu} - g^{\mu\nu}{}_{,\nu} = 0$, there is no curvature, $R^\alpha{}_{\beta\mu\nu} = 0$. If this can be carried over into momentum space through a suitable definition of Christoffel (affine)-type connections $\Gamma^\mu{}_{\alpha\beta}$ and a curvature tensor $R^\alpha{}_{\beta\mu\nu}$, and if we can do this in a way that retains the experimental phenomenology of the magnetic moment anomaly without compromise, then Figure 2 would mean that in quantum field theory, *interaction vertexes are proportional to the geometric curvature of momentum space*, that is, "interaction $\propto$ curvature." If sustainable, this would place quantum field theory onto a firm, general relativistic footing, with the Ward-Takahashi identity standing at the center of this relationship. This provides a strong motivation to carefully consider the $g_{\mu\nu} = \Gamma_{\mu\nu}$ hypothesis.

## 4. Christoffel Connections and Riemann Curvature in Momentum Space; Non-Longitudinal Takahashi Terms

We first *define* a set of Christoffel (affine)-type connections in momentum space in the same manner as in spacetime:

$$\Gamma^\mu{}_{\alpha\beta}(p,q,p') \equiv \tfrac{1}{2} g^{\mu\nu} \left( g_{\nu\alpha/\beta} + g_{\beta\nu/\alpha} - g_{\alpha\beta/\nu} \right). \tag{4.1}$$

So too, we define covariant differentiation as in spacetime. So, $A_{\mu);\alpha}(p,q,p') \equiv A_{\mu/\alpha} - \Gamma^\sigma{}_{\mu\alpha} A_\sigma$, for example. And, for $g_{\mu\nu}$, (4.1) ensures, by identity, that:

$$g_{\mu\nu}(p,q,p')_{\rangle\alpha} \equiv g_{\mu\nu/\alpha} - \Gamma^{\sigma}{}_{\mu\alpha} g_{\sigma\nu} - \Gamma^{\sigma}{}_{\alpha\nu} g_{\mu\sigma} = 0. \tag{4.2}$$

Now, from (3.4), we may calculate the ordinary derivative of $g_{\mu\nu}$ to be:

$$g_{\mu\nu}(p,q,p')_{/\alpha} = -F_2 \Gamma_\nu \left( p_{\mu/\alpha} + \tfrac{1}{2} q_{\mu/\alpha} \right) - F_2 \Gamma_\mu \left( p_{\nu/\alpha} + \tfrac{1}{2} q_{\nu/\alpha} \right). \tag{4.3}$$

From (3.4) and (4.1), we may also calculate the connections (note $p_{\alpha/\beta} = p_{\beta/\alpha} = g_{\alpha\beta}$):

$$\begin{aligned}\Gamma^{\mu}{}_{\alpha\beta}(p,q,p') &= +\tfrac{1}{4} F_2 \Gamma^{\mu} \left( +2 p_{\alpha/\beta} + 2 p_{\beta/\alpha} + q_{\alpha/\beta} + q_{\beta/\alpha} \right) \\ &\quad + \tfrac{1}{4} F_2 \Gamma_\alpha g^{\mu\nu} \left( q_{\nu/\beta} - q_{\beta/\nu} \right) + \tfrac{1}{4} F_2 \Gamma_\beta g^{\mu\nu} \left( q_{\nu/\alpha} - q_{\alpha/\nu} \right)\end{aligned} \tag{4.4}$$

Using (4.2) and (4.4), we may then calculate for $(p,q,p')$, that:

$$\Gamma^{\sigma}{}_{\mu\alpha} g_{\sigma\nu} + \Gamma^{\sigma}{}_{\alpha\nu} g_{\mu\sigma} = -F_2 \Gamma_\nu \left( p_{\mu/\alpha} + \tfrac{1}{2} q_{\mu/\alpha} \right) - F_2 \Gamma_\mu \left( p_{\nu/\alpha} + \tfrac{1}{2} q_{\nu/\alpha} \right) = g_{\mu\nu/\alpha}. \tag{4.5}$$

This confirms that $g_{\mu\nu}(p,q,p')_{\rangle\alpha} = 0$ and serves as a check on (4.4). The contraction of (4.3), using $p_\mu{}^{/\nu} = \delta_\mu{}^\nu$ and $p_\nu{}^{/\nu} = 1$, yields:

$$g_{\mu\nu}(p,q,p')^{/\nu} = -2 F_2 \left( \Gamma_\mu + \tfrac{1}{4} \Gamma_\nu q_\mu{}^{/\nu} + \tfrac{1}{4} \Gamma_\mu q_\nu{}^{/\nu} \right) \quad \left( = \Gamma^{\alpha}{}_{\mu\alpha} + g^{\nu\alpha} g_{\mu\sigma} \Gamma^{\sigma}{}_{\alpha\nu} \right). \tag{4.6}$$

It is also helpful to be aware of:

$$\Gamma^{\alpha}{}_{\mu\alpha}(p,q,p') = \sqrt{-g}_{/\mu} \Big/ \sqrt{-g} = F_2 \left( \Gamma_\mu + \tfrac{1}{2} \Gamma^{\alpha} q_{\alpha/\mu} \right). \tag{4.7}$$

Now, we employ (4.6) in the Ward-Takahashi identities (3.11), (3.12). Equation (3.11) for Takahashi now becomes:

$$\begin{aligned}q^\mu g_{\mu\nu}(p,q,p')^{\rangle\nu} = 0 &= q^\mu g_{\mu\nu}(p,q,p')^{/\nu} + 2 F_2 \left( \Gamma_\mu q^\mu + \tfrac{1}{4} \Gamma_\nu q^\mu q_\mu{}^{/\nu} + \tfrac{1}{4} \Gamma_\mu q^\mu q_\nu{}^{/\nu} \right) \\ &= -2 F_2 \left( \Gamma_\mu q^\mu + \tfrac{1}{4} \Gamma_\nu q^\mu q_\mu{}^{/\nu} + \tfrac{1}{4} \Gamma_\mu q^\mu q_\nu{}^{/\nu} \right) + 2 F_2 \left( S'_F{}^{-1}(p') - S'_F{}^{-1}(p) \right).\end{aligned} \tag{4.8}$$

For $q^\mu \to 0$, this reduces, unmodified, to the Ward identity (3.12):

$$g_{\mu\nu}(p,0,p)^{\rangle\nu} = 0 = g_{\mu\nu}(p,0,p)^{/\nu} + 2 F_2 \Gamma_\mu = -2 F_2 \Gamma_\mu + 2 F_2 S'_F{}^{-1}{}_{/\mu}. \tag{4.9}$$

Because the Ward identity emerges naturally from the definition of covariant differentiation in momentum space, this means, when other vectors and tensors are covariantly differentiated in the usual way using the momentum space Christoffel-type connections $\Gamma^{\mu}{}_{\alpha\beta}$, that we are implicitly using the Ward identity to drive the differentiation, and that this identity will therefore be satisfied whenever we take a covariant derivative. This may help achive renormalizablity.

What is also of interest is that for $(p, q, p')$ generally, the Takahashi generalization (4.8) obtains extra terms in addition to the usual longitudinal $\Gamma_\mu q^\mu$, when it is obtained through covariant differentiation in momentum space. Renaming dummy indexes in (4.8), we find:

$$q^\mu g_{\mu\nu}(p,q,p')^{)\nu} - q^\mu g_{\mu\nu}(p,q,p')^{/\nu} = 2F_2\left(q^\mu + \tfrac{1}{4}q^\nu q_\nu{}^{/\mu} + \tfrac{1}{4}q^\mu q_\nu{}^{/\nu}\right)\Gamma_\mu$$
$$= 2F_2\left(S'{}_F^{-1}(p') - S'{}_F^{-1}(p)\right) \quad \left(= -q^\mu \Gamma^\alpha{}_{\mu\alpha} - g^{\nu\alpha} q_\sigma \Gamma^\sigma{}_{\alpha\nu}\right) \tag{4.10}$$

This leads to a modification of Figures 1 and 2, into the form:

$$q^\mu\left[g_{\mu\nu}(p,q,p')^{)\nu} - g_{\mu\nu}(p,q,p')^{/\nu}\right] = 2F_2\left(q^\mu + \tfrac{1}{4}q^\nu q_\nu{}^{/\mu} + \tfrac{1}{4}q^\mu q_\nu{}^{/\nu}\right)$$

$$= 2F_2\left[\underset{p+q}{\circ}\underset{p+q^{-1}}{\phantom{o}} - \underset{p}{\circ}\underset{p^{-1}}{\phantom{o}}\right]$$

Figure 3

It is possible that the extra terms $+\tfrac{1}{4}q^\nu q_\nu{}^{/\mu} + \tfrac{1}{4}q^\mu q_\nu{}^{/\nu}$ multiplying the vertex can help to better develop transverse generalizations of Takahashi. For example, from (4.8), using $q^\mu q_\mu{}^{/\nu} = g_{\mu\sigma}{}^{/\nu} q^\mu q^\sigma + q_\mu q^{\mu/\nu}$; $q^\mu q_\nu{}^{/\nu} = (q^\mu q_\nu)^{/\nu} - q^{\mu/\nu} q_\nu$, and $q^\sigma q_\sigma = 0$ for a massless on-shell photon in $(g_{\mu\sigma} q^\mu q^\sigma)^{/\nu} = g_{\mu\sigma}{}^{/\nu} q^\mu q^\sigma + g_{\mu\sigma}(q^\mu q^\sigma)^{/\nu}$, it is possible to write:

$$S'{}_F^{-1}(p') - S'{}_F^{-1}(p) = \Gamma_\mu q^\mu + \tfrac{1}{4}\Gamma_\nu q^\mu q_\mu{}^{/\nu} + \tfrac{1}{4}\Gamma_\mu q^\mu q_\nu{}^{/\nu}$$
$$= q^\mu \Gamma_\mu + \tfrac{1}{4}q^{\mu/\nu}(q_\mu \Gamma_\nu - q_\nu \Gamma_\mu) + \tfrac{1}{4}\left[(q^\mu q_\nu)^{/\nu}\Gamma_\mu - g_{\mu\sigma}(q^\mu q^\sigma)^{/\nu}\Gamma_\nu\right] \tag{4.11}$$

Determining the transverse part of the vertex function is to date an unresolved problem, and many different *ad hoc* approachs have been taken. [6],[7],[8] In the approach leading to (4.11), the *ansatz* is provided by the definition of covariant differentiation in momentum space, and specific terms arise from the contraction $\Gamma^\alpha{}_{\mu\alpha} + g^{\nu\alpha} g_{\mu\sigma} \Gamma^\sigma{}_{\alpha\nu}$ of the covariant derivative terms $\Gamma^\sigma{}_{\mu\alpha} g_{\sigma\nu} + \Gamma^\sigma{}_{\alpha\nu} g_{\mu\sigma}$ see (4.6), (4.2). In (4.11) this yields a longitudinal (divergence) term of the form $\Gamma_\mu q^\mu$, a transverse (curl) term of the form $q^{\mu/\nu}(q_\mu \Gamma_\nu - q_\nu \Gamma_\mu)$, and an additional term of the form $(q^\mu q_\nu)^{/\nu}\Gamma_\mu - g_{\mu\sigma}(q^\mu q^\sigma)^{/\nu}\Gamma_\nu$.[9] Yet, these terms are all contracted to scalars, so the propagator form $S'{}_F^{-1}(p') - S'{}_F^{-1}(p)$ remains intact and it is unnecessary to construct new forms such as the commonly-used $\sigma^{\mu\nu} S'{}_F^{-1}(p') + S'{}_F^{-1}(p) \sigma^{\mu\nu}$, or forms involving $\gamma^5$, for example. [6], [7], [8], [9]

Having established these simple, yet fundamental formal connections, it becomes critical to show that these results do not conflict with any experimental results, starting with the anomalous magnetic moment. That is, we need to show that a Lagrangian density with a metric tensor $g_{\mu\nu} = \Gamma_{\mu\nu}$, and $\Gamma^\mu$ in place of $\gamma^\mu$ at all vertexes, can describe the magnetic anomaly with equal facility as a Lagrangian density with $g_{\mu\nu} = \eta_{\mu\nu}$, and with $\Gamma^\mu$ only at the $\overline{e\psi}\Gamma^\mu A_\mu \psi$ vertex.

Before moving to this next task, using $\Gamma^\mu{}_{\alpha\beta}(p,0,p) = F_2\,\Gamma^\mu(p,0,p)p_{\alpha/\beta}$ from (4.4), we pause to calculate the momentum space Riemann tensor $R^\alpha{}_{\beta\mu\nu}(p,0,p)$. Because the Feynman graph ⎯○⎯ $^{p\ \ \ p}$ is equivalent to Figure 2 with $q=0$, this means that $g^{\mu\nu}(p,0,p)_{\rangle\nu} = g^{\mu\nu}(p,0,p)_{/\nu}$. Since, these covariant and ordinary derivatives are equal, we anticipate that $R^\alpha{}_{\beta\mu\nu}(p,0,p) = 0$. This can serve as a check on everything derived so far, so let's see if this is so.

As in spacetime, we *define* the Riemann tensor $R^\alpha{}_{\beta\mu\nu}$ in momentum space as a measure of the degree to which covariant derivatives do not commute, i.e., $R^\alpha{}_{\beta\mu\nu}A_\alpha \equiv A_{\beta\rangle\mu\rangle\nu} - A_{\beta\rangle\nu\rangle\mu}$. We also note from (4.4), that because the Christoffel-type connections $\Gamma^\mu{}_{\alpha\beta}$ contain $\Gamma^\mu$, in general, they will *not* commute. Thus, for $R^\alpha{}_{\beta\mu\nu}$ in momentum space, we write:

$$R^\alpha{}_{\beta\mu\nu} = -\Gamma^\alpha{}_{\beta\mu/\nu} + \Gamma^\alpha{}_{\beta\nu/\mu} + \tfrac{1}{2}\left(\Gamma^\sigma{}_{\beta\nu}\Gamma^\alpha{}_{\sigma\mu} + \Gamma^\alpha{}_{\sigma\mu}\Gamma^\sigma{}_{\beta\nu}\right) - \tfrac{1}{2}\left(\Gamma^\sigma{}_{\beta\mu}\Gamma^\alpha{}_{\sigma\nu} + \Gamma^\alpha{}_{\sigma\nu}\Gamma^\sigma{}_{\beta\mu}\right). \tag{4.12}$$

If the $\Gamma^\mu{}_{\alpha\beta}$ *do* commute as in spacetime, then (4.12) is identical to the usual definition of $R^\alpha{}_{\beta\mu\nu}$. However, if the $\Gamma^\mu{}_{\alpha\beta}$ do *not* commute, the usual $R^\alpha{}_{\beta\mu\nu}$ expression, when contracted, leads to the Ricci tensor $R_{\beta\mu}$ being non-symmetric, $R_{\beta\mu} \neq R_{\mu\beta}$. If we impose the *requirement* that $R_{\beta\mu} = R_{\mu\beta}$, then it is necessary to employ $+\tfrac{1}{2}\left(\Gamma^\sigma{}_{\beta\nu}\Gamma^\alpha{}_{\sigma\mu} + \Gamma^\alpha{}_{\sigma\mu}\Gamma^\sigma{}_{\beta\nu}\right) - \tfrac{1}{2}\left(\Gamma^\sigma{}_{\beta\mu}\Gamma^\alpha{}_{\sigma\nu} + \Gamma^\alpha{}_{\sigma\nu}\Gamma^\sigma{}_{\beta\mu}\right)$ to achieve $R_{\beta\mu} = R_{\mu\beta}$ in the face of non-commuting $\Gamma^\mu{}_{\alpha\beta}$.

Now, we calculate. Substituting $\Gamma^\mu{}_{\alpha\beta}(p,0,p)$ from (4.4) into (4.12) first yields:

$$\begin{aligned}R^\alpha{}_{\beta\mu\nu} &= -F_2\left(\Gamma^\alpha{}_{/\nu}p_{\beta/\mu} + \Gamma^\alpha p_{\beta/\mu/\nu}\right) + F_2\left(\Gamma^\alpha{}_{/\mu}p_{\beta/\nu} + \Gamma^\alpha p_{\beta/\nu/\mu}\right) \\ &+ \tfrac{1}{2}F_2{}^2\left(\Gamma^\sigma\Gamma^\alpha p_{\beta/\nu}p_{\sigma/\mu} + \Gamma^\alpha\Gamma^\sigma p_{\sigma/\mu}p_{\beta/\nu}\right) - \tfrac{1}{2}F_2{}^2\left(\Gamma^\sigma\Gamma^\alpha p_{\beta/\mu}p_{\sigma/\nu} + \Gamma^\alpha\Gamma^\sigma p_{\sigma/\nu}p_{\beta/\mu}\right)\end{aligned}. \tag{4.13}$$

Because the *ordinary* derivatives commute, $p_{\beta/\mu/\nu} - p_{\beta/\nu/\mu} = 0$, this reduces to:

$$\begin{aligned}R^\alpha{}_{\beta\mu\nu} &= F_2\left(-\Gamma^\alpha{}_{/\nu}p_{\beta/\mu} + \Gamma^\alpha{}_{/\mu}p_{\beta/\nu}\right) + \tfrac{1}{2}F_2{}^2\{\Gamma^\sigma\Gamma^\alpha + \Gamma^\alpha\Gamma^\sigma\}\left(p_{\beta/\nu}p_{\sigma/\mu} - p_{\beta/\mu}p_{\sigma/\nu}\right) \\ &= F_2\left(-\Gamma^\alpha{}_{/\nu}p_{\beta/\mu} + \Gamma^\alpha{}_{/\mu}p_{\beta/\nu}\right) + F_2{}^2 g^{\sigma\alpha}\left(p_{\beta/\nu}p_{\sigma/\mu} - p_{\beta/\mu}p_{\sigma/\nu}\right)\end{aligned}. \tag{4.14}$$

where in the second line we have employed $\Gamma^{\sigma\alpha} = \tfrac{1}{2}F_2{}^2\{\Gamma^\sigma\Gamma^\alpha + \Gamma^\alpha\Gamma^\sigma\}$, the anticommutator (2.7), with $\Gamma^{\sigma\alpha} = g^{\sigma\alpha}$. Now, all with $q^\mu = 0$, we substitute $\Gamma^\alpha = (F_1 + F_2 m)\gamma^\alpha - F_2 p^\alpha$ from (3.3), we observe therefore that $\Gamma^\alpha{}_{/\nu} = -F_2 p^\alpha{}_{/\nu}$, and we make use of $p^\alpha{}_{/\nu} = \delta^\alpha{}_\nu$, $p_{\sigma/\nu} = g_{\sigma\nu}$ and $\delta^\alpha{}_\nu = g^{\sigma\alpha}g_{\sigma\nu}$, to reduce to:

$$\begin{aligned}
R^{\alpha}{}_{\beta\mu\nu} &= F_2^2 \left(p^{\alpha}{}_{/\nu} p_{\beta/\mu} - p^{\alpha}{}_{/\mu} p_{\beta/\nu}\right) + F_2^{\ 2} g^{\sigma\alpha}\left(p_{\beta/\nu} p_{\sigma/\mu} - p_{\beta/\mu} p_{\sigma/\nu}\right) \\
&= F_2^2 \left(\left(p^{\alpha}{}_{/\nu} - g^{\sigma\alpha} p_{\sigma/\nu}\right) p_{\beta/\mu} - \left(p^{\alpha}{}_{/\mu} - g^{\sigma\alpha} p_{\sigma/\mu}\right) p_{\beta/\nu}\right) \\
&= F_2^2 \left(\left(\delta^{\alpha}{}_{\nu} - g^{\sigma\alpha} g_{\sigma\nu}\right) p_{\beta/\mu} - \left(\delta^{\alpha}{}_{\mu} - g^{\sigma\alpha} g_{\sigma\mu}\right) p_{\beta/\nu}\right) \\
&= F_2^2 \left(\left(\delta^{\alpha}{}_{\nu} - \delta^{\alpha}{}_{\nu}\right) p_{\beta/\mu} - \left(\delta^{\alpha}{}_{\mu} - \delta^{\alpha}{}_{\mu}\right) p_{\beta/\nu}\right) = 0
\end{aligned} \qquad (4.15)$$

We do indeed find that $R^{\alpha}{}_{\beta\mu\nu}(p,0,p) = 0$, that is, that the $(p,0,p)$ limit for a non-interacting fermion line —p—○—p— describes a zero curvature in momentum space.

Because $R^{\alpha}{}_{\beta\mu\nu}(p,0,p) = 0$, and because the metric tensor is identical to the Minkowski tensor, $g^{\mu\nu} = \eta^{\mu\nu}$ whenever *all* components of the Riemann tensor are equal to zero as here, this means that the $g^{\mu\nu}(p,0,p)$ derived from (3.4) must be equal to the Minkowski metric, that is:

$$g^{\mu\nu}(p,0,p) = (F_1 + F_2 m)^2 \eta^{\mu\nu} - (F_1 + F_2 m) F_2 \left(\gamma^{\mu} p^{\nu} + \gamma^{\nu} p^{\mu}\right) + F_2^2 p^{\mu} p^{\nu} = \eta^{\mu\nu} \qquad (4.16)$$

As we shall see in section 8, this equation has utility in the form of a Dirac-type field equation operating on a fermion wavefunction, $\left(g^{\mu\nu}(p,0,p) - \eta^{\mu\nu}\right)\psi = 0$.

One can, and eventually should, calculate the curvature tensor $R^{\alpha}{}_{\beta\mu\nu}(p',q,p) \neq 0$ from $\Gamma^{\mu}{}_{\alpha\beta}(p',q,p)$ in (4.4), thereby generalizing (4.15), both of which generalize —p—○—p— to the full interaction vertex for Figure 2, with $q \neq 0$. We leave this, for now, to a future paper, and to the interested reader.

All of above, of course, is based on *hypothesizing* that $\Gamma_{\mu\nu} = g_{\mu\nu}$. We now examine if this hypothesis, which leads to $\mathcal{L}' = i\bar{\psi}\Gamma^{\mu}\Gamma_{\mu\nu}\partial^{\nu}\psi - e'\bar{\psi}\Gamma^{\mu}\Gamma_{\mu\nu}A^{\nu}\psi - m'\bar{\psi}\psi - \frac{1}{4}F^{\mu\nu}F_{\mu\nu}$, can be reconciled with the usual $\mathcal{L} = i\bar{\psi}\gamma^{\mu}\eta_{\mu\nu}\partial^{\nu}\psi + e\bar{\psi}\Gamma^{\mu}\eta_{\mu\nu}A^{\nu}\psi - m\bar{\psi}\psi - \frac{1}{4}F^{\mu\nu}F_{\mu\nu}$, and, most importantly, can be used to explain the observed magnetic moment anomaly. After all, none of the formal results above are pertinent if the magnetic anomaly cannot be explained equally-well using $\mathcal{L}$ or using $\mathcal{L}'$. If, conversely, one *can* explain the magnetic anomaly on the basis of $\mathcal{L}'$ just as well as $\mathcal{L}$, then the results derived here, at least insofar as the magnetic anomaly is concerned, are not falsified by experimental observation. This crucial examination, now carried out in detail, will be the main subject of sections 5 through 7.

## 5. The Classical Hamiltonian and the Magnetic Anomaly, in the Customary Derivation

In this section, we examine the magnetic anomaly utilizing the customary Dirac Lagrangian density $\mathcal{L} = i\bar{\psi}\gamma^{\mu}\eta_{\mu\nu}\partial^{\nu}\psi + e\bar{\psi}\Gamma^{\mu}\eta_{\mu\nu}A^{\nu}\psi - m\bar{\psi}\psi - \frac{1}{4}F^{\mu\nu}F_{\mu\nu}$. This is to establish a baseline for later comparison. In section 6, we will derive the anticommutators $\Gamma_{\mu\nu}$ for a photon, which is a necessary intermediate result. In section 7, we will see how to reconcile the Lagrangian density $\mathcal{L}$ with $\mathcal{L}' = i\bar{\psi}\Gamma^{\mu}\Gamma_{\mu\nu}\partial^{\nu}\psi - e'\bar{\psi}\Gamma^{\mu}\Gamma_{\mu\nu}A^{\nu}\psi - m'\bar{\psi}\psi - \frac{1}{4}F^{\mu\nu}F_{\mu\nu}$, in the $A^{\mu} = 0$, $q^{\mu} = 0$ limit, using the $\Gamma_{\mu\nu}$ to be derived in section 6 as the covariant metric tensor. That is, we show how to reconcile $\mathcal{L}$ with a Lagrangian $\mathcal{L}'$ in which $\Gamma_{\mu\nu} = \frac{1}{2}\{\Gamma_{\mu}, \Gamma_{\nu}\} = g_{\mu\nu}$ is employed as the metric tensor and $\Gamma^\mu$ is employed at all vertexes, and in which the only difference between $\mathcal{L}$ and $\mathcal{L}'$ is that there subsists a particular set of relationships between $e$ and $e'$, $m$ and $m'$, and $p^\mu$ and $p^{\mu'}$, such that $\mathcal{L}(m, e, p^\mu) = \mathcal{L}'(m', e', p^{\mu'})$.

To begin with, we briefly review how $\mathcal{L}$ is renormalized. The total *bare* QED Lagrangian density, absorbing the *infinite*, divergent quantities, and including the *finite*, convergent contribution $\Lambda^\mu$ to the fermion-photon vertex via the substitution $\gamma^\mu \Rightarrow \Gamma^\mu = \gamma^\mu + \Lambda^\mu$, is (see [5], equation (9.132)):

$$\mathcal{L}_B = iZ_2\bar{\psi}\gamma^\mu\partial_\mu\psi + e\mu^{\varepsilon/2}Z_1\bar{\psi}\Gamma^\mu A_\mu\psi - (m+A)\bar{\psi}\psi - \tfrac{1}{4}Z_3 F^{\mu\nu}F_{\mu\nu} - \tfrac{1}{2}Z_3 A^\mu{}_{;\mu}A^\nu{}_{;\nu}. \tag{5.1}$$

All of the bare objects implicit in the above are renormalized using "infinite" constants according to (see [5], equations (9.111), (9.134), (9.125), (9.113)):

$$\psi_B = Z_2^{.5}\psi; \quad e_B = e\mu^{\varepsilon/2}\frac{Z_1}{Z_2}Z_3^{-.5} = e\mu^{\varepsilon/2}Z_3^{-.5}; \quad A_B{}^\mu = Z_3^{.5}A^\mu; \quad m_B = Z_2^{-1}(m+A) = m + \delta m, \tag{5.2}$$

where the Ward identity considered in the previous sections leads to $Z_1 = Z_2$, [5], equation (7.129). For one-loop QED renormalization, we have ([5], equation (9.133)):

$$Z_1 = Z_2 = 1 - \frac{e^2}{8\pi^2\varepsilon} = 1 - \frac{\alpha}{2\pi\varepsilon}; \quad Z_3 = 1 - \frac{e^2}{6\pi^2\varepsilon} = 1 - \frac{2\alpha}{3\pi\varepsilon}; \quad A = -\frac{me^2}{2\pi^2\varepsilon} = -\frac{2m\alpha}{\pi\varepsilon}, \tag{5.3}$$

where $\varepsilon \to 0$ drives each of these constants to infinity, while the Ward identity ensures renormalizability to all orders. Combining (5.1) and (5.2) leads readily to:

$$\mathcal{L}_B = i\bar{\psi}_B\gamma^\mu\partial_\mu\psi_B + e_B\bar{\psi}_B\Gamma^\mu A_{\mu B}\psi_B - m_B\bar{\psi}_B\psi_B - \tfrac{1}{4}F^{\mu\nu}{}_B F_{\mu\nu B} - \tfrac{1}{2}A^\mu{}_{;\mu}A^\nu{}_{;\nu}. \tag{5.4}$$

The renormalizability of the above allows us to represent the observed, dressed, physical Lagrangian as:

$$\mathcal{L} = i\bar{\psi}\gamma^\mu\partial_\mu\psi + e\bar{\psi}\Gamma^\mu A_\mu\psi - m\bar{\psi}\psi - \tfrac{1}{4}F^{\mu\nu}F_{\mu\nu} - \tfrac{1}{2}A^\mu{}_{;\mu}A^\nu{}_{;\nu}. \tag{5.5}$$

where the physical value of each of the quantities above is arrived at using (5.2) to absorb the logarithmically-divergent infinities of the bare quantities.

We shall now examine this customary Lagrangian (5.5), using the corresponding Dirac equation in momentum space, in the form $(\gamma^\mu g_{\mu\nu}p^\nu - e\Gamma^\mu g_{\mu\nu}A^\nu - m)u(p) = 0$, with $g_{\mu\nu} = \eta_{\mu\nu}$. It is very important to observe four things here: First, the metric tensor is regarded to be $g_{\mu\nu} = \eta_{\mu\nu}$ because gravitation, at least as it is understood at present, is presumed to be something which may be entirely neglected when considering, say, the interaction of a fermion field quantum (e.g., electron) with an electromagnetic field quantum (i.e., photon) at observable energies. Thus, it is $\eta_{\mu\nu}$ which is used to raise and lower spacetime indexes and generally to couple objects together in spacetime. Second, the finite perturbative correction $\Lambda^\mu$ in $\Gamma^\mu = \gamma^\mu + \Lambda^\mu$ is employed in the electron / photon vertex term $e\bar{\psi}\Gamma^\mu \eta_{\mu\nu} A^\nu \psi$, *but only in this term*. Third, in contrast, the kinetic energy term $i\bar{\psi}\gamma^\mu \eta_{\mu\nu} \partial^\nu \psi$ employs the bare vertex $\gamma^\mu$ and *not* the perturbative vertex $\Gamma^\mu = \gamma^\mu + \Lambda^\mu$. Fourth, *nowhere is the anticommutator $\Gamma_{\mu\nu}$ employed*.

Substituting $g_{\mu\nu} = \eta_{\mu\nu}$ and (3.1) with photon momentum $q^\mu \equiv (p'-p)^\mu$ hence $(p'+p)^\mu = 2p^\mu + q^\mu$ into $(\gamma^\mu g_{\mu\nu} p^\nu - e\Gamma^\mu g_{\mu\nu} A^\nu - m)u(p) = 0$, yields:

$$(\gamma^\mu \eta_{\mu\nu} p^\nu - e((F_1 + F_2 m)\gamma^\mu - F_2(p^\mu + \tfrac{1}{2} q^\mu))\eta_{\mu\nu} A^\nu - m)u(p) = 0. \tag{5.6}$$

This is equivalent to the paired equations in the Dirac representation:

$$\begin{cases} [E - (F_1 + F_2 m)e\phi + F_2((E + \tfrac{1}{2}\omega)e\phi - (\mathbf{p} + \tfrac{1}{2}\mathbf{q}) \cdot e\mathbf{A}) - m]u - \sigma \cdot (\mathbf{p} - (F_1 + F_2 m)e\mathbf{A})v = 0 \\ \sigma \cdot (\mathbf{p} - (F_1 + F_2 m)e\mathbf{A})u - [E - (F_1 + F_2 m)e\phi - F_2((E + \tfrac{1}{2}\omega)e\phi + (\mathbf{p} + \tfrac{1}{2}\mathbf{q}) \cdot e\mathbf{A}) + m]v = 0 \end{cases}. \tag{5.7}$$

Here, $\sigma$ are the Pauli spin matrices, the photon momentum $q^\mu = (\omega, q_x, q_y, q_z)$, the electromagnetic field potential $A^\mu = (\phi, A_x, A_y, A_z)$, the energy-momentum vector for the fermion is $p^\mu = (E, p_x, p_y, p_z)$, $u(p)$ is the two-component particle spinor, $v(p)$ the two-component antiparticle spinor, and $m$ is the rest mass of the fermion. These are combined by isolating $v(p)$ and inserting the resulting expression into the equation for $u(p)$, to yield the classical Hamiltonian $H$:

$$\begin{aligned} H = E - m &= (F_1 + F_2 m)e\phi - F_2[(E + \tfrac{1}{2}\omega)e\phi - (\mathbf{p} + \tfrac{1}{2}\mathbf{q}) \cdot e\mathbf{A}] \\ &+ \frac{\sigma \cdot (\mathbf{p} - (F_1 + F_2 m)e\mathbf{A})\sigma \cdot (\mathbf{p} - (F_1 + F_2 m)e\mathbf{A})}{E + m - (F_1 + F_2 m)e\phi - F_2((E + \tfrac{1}{2}\omega)e\phi + (\mathbf{p} + \tfrac{1}{2}\mathbf{q}) \cdot e\mathbf{A})}. \end{aligned} \tag{5.8}$$

From there, we reduce as usual, based on the relationship $\sigma^i \sigma^j = \delta^{ij} + i\varepsilon^{ijk}\sigma^k$, the quantum mechanical substitution $p^\mu \to i\hbar \partial^\mu$ and the electromagnetic field $F^{\mu\nu} = \partial^\mu A^\nu - \partial^\nu A^\mu$, applied in the form:

$$\begin{aligned} &\sigma \cdot (\mathbf{p} - (F_1 + F_2 m)e\mathbf{A})\sigma \cdot (\mathbf{p} - (F_1 + F_2 m)e\mathbf{A}) \\ &= (\mathbf{p} - (F_1 + F_2 m)e\mathbf{A})^2 - ie\sigma \cdot (F_1 + F_2 m)(\mathbf{p} \times \mathbf{A} + \mathbf{A} \times \mathbf{p}), \\ &= (\mathbf{p} - (F_1 + F_2 m)e\mathbf{A})^2 - 2e\hbar(F_1 + F_2 m)\frac{\sigma}{2} \cdot \mathbf{B} \end{aligned} \tag{5.9}$$

to yield:

$$H = E - m = +(F_1 + F_2 m)e\phi - F_2\left((E + \tfrac{1}{2}\omega)e\phi + (\mathbf{p} + \tfrac{1}{2}\mathbf{q})\cdot e\mathbf{A}\right)$$
$$+ \frac{(\mathbf{p} - (F_1 + F_2 m)e\mathbf{A})^2}{E + m - (F_1 + F_2 m)e\phi - F_2\left((E + \tfrac{1}{2}\omega)e\phi + (\mathbf{p} + \tfrac{1}{2}\mathbf{q})\cdot e\mathbf{A}\right)} \quad (5.10)$$
$$- \frac{4m(F_1 + F_2 m)}{E + m - (F_1 + F_2 m)e\phi - F_2\left((E + \tfrac{1}{2}\omega)e\phi + (\mathbf{p} + \tfrac{1}{2}\mathbf{q})\cdot e\mathbf{A}\right)} \frac{e\hbar}{2m}\frac{\sigma}{2}\cdot \mathbf{B}$$

The term on the final line multiplying $-\frac{e\hbar}{2m}\frac{\sigma}{2}\cdot \mathbf{B}$ is the gyromagnetic ratio $g$, i.e.:

$$\frac{g}{2} = \frac{2m(F_1 + F_2 m)}{E + m - (F_1 + F_2 m)e\phi - F_2\left((E + \tfrac{1}{2}\omega)e\phi + (\mathbf{p} + \tfrac{1}{2}\mathbf{q})\cdot e\mathbf{A}\right)}. \quad (5.11)$$

For a fermion taken in its rest frame $E = m$, with $A^\mu = 0$, (5.10) becomes:

$$H = E - m = -\frac{4m(F_1 + F_2 m)}{E + m}\frac{e\hbar}{2m}\frac{\sigma}{2}\cdot \mathbf{B} \stackrel{E=m}{=} -2(F_1 + F_2 m)\frac{e\hbar}{2m}\frac{\sigma}{2}\cdot \mathbf{B} = -g\frac{e\hbar}{2m}\frac{\sigma}{2}\cdot \mathbf{B}. \quad (5.12)$$

Using the Schwinger form factors (2.5), this reduces for $E = m$, as expected, to:

$$H = E - m = -2\left(1 + \frac{\alpha}{2\pi}\right)\frac{e\hbar}{2m}\frac{\sigma}{2}\cdot \mathbf{B} = -g\frac{e\hbar}{2m}\frac{\sigma}{2}\cdot \mathbf{B}. \quad (5.13)$$

i.e.,

$$\frac{g}{2} = F_1 + mF_2 = 1 + \frac{\alpha}{2\pi}. \quad (5.14)$$

In all of the above, which is well-known and describes what is experimentally observed, the rest mass $m$ and energy component $E = p^0$ are understood to represent the *observed, renormalized* rest mass and energy of the fermion, $e$ and $\alpha$ are understood to represent the *observed, renormalized* charge and coupling strengths, and $g$ is understood to be the observed gyromagnetic ratio with magnitude determined by the renormalized $\alpha$. This is in contrast to all of these quantities being in the "bare" form of (5.4). If one desires to know, for example, how the observed charge runs as a function of probe energy, one employs $e = Z_3^{.5}\mu^{-\varepsilon/2}e_B$ from (5.2). At the one loop approximation, using $Z_3^{.5} \cong 1 - \frac{e^2}{12\pi^2\varepsilon}$ from (5.3), $e\left(1 + \frac{e^2}{12\pi^2\varepsilon}\right) \cong \mu^{-\varepsilon/2}e_B$ which, for $\varepsilon \to 0$, yields the differential equation $\mu\frac{\partial e}{\partial \mu} \cong \frac{e^3}{12\pi^2}$. The solution to this is the very familiar running relationship $e^2(\mu) = e^2(\mu_0)\left/\left(1 - \frac{e^2(\mu_0)}{6\pi^2}\ln\frac{\mu}{\mu_0}\right)\right.$. (See [5], pp. 345-347.)

## 6. Exact Anticommutators for a Massless Vector Boson

At this point, we derive the anticommutators $\Gamma_{\mu\nu}$ for a massless vector boson. This is interesting in its own right, but we derive these at this time is because, when we calculate the Hamiltonian $H'$ for $\mathcal{L}'$ in section 7 to contrast with $H$ for $\mathcal{L}$ in section 5, we will need to employ these $\Gamma_{\mu\nu}$ as the covariant metric tensor $g_{\mu\nu} = \Gamma_{\mu\nu}$ to form couplings in spacetime. So, the main purpose of this section is to derive the covariant metric tensor to be used in section 7.

We start by considering the anticommutators (2.11) for a photon. Setting $q = p'-p$ and $q^\sigma \equiv (p'-p)^\sigma$, we write:

$$\Gamma_{\mu\nu} = F_1^2 \eta_{\mu\nu} - \tfrac{1}{4} F_2^2 \left[ \eta_{\mu\nu} q^2 - \eta_{\sigma\mu} \eta_{\nu\tau} q^\sigma q^\tau \right]. \tag{6.1}$$

Now, consider the above for a massless on-shell photon with frequency $\omega$ moving along the z-axis in Cartesian coordinates, i.e., $q^\sigma = (\omega,0,0,\omega)$. Noting from (6.1) that the off-diagonal components $\Gamma_{03} = \Gamma_{30}$ will be non-zero by virtue of the $\eta_{\sigma\mu}\eta_{\nu\tau}q^\sigma q^\tau$ term, first, we may write:

$$\slashed{q} = \gamma^\mu q_\mu = \gamma^\mu \Gamma_{\mu\nu} q^\nu = \left( \gamma^0 \Gamma_{00} + \gamma^3 \Gamma_{33} + \gamma^0 \Gamma_{03} + \gamma^3 \Gamma_{30} \right) \omega . \tag{6.2}$$

Next, we use (6.1) to deduce the pertinent $\Gamma_{\mu\nu}$ needed in (6.2). Thus:

$$\Gamma_{00} = F_1^2 - \tfrac{1}{4} F_2^2 \left[ q^2 - \omega^2 \right]; \quad \Gamma_{33} = -F_1^2 + \tfrac{1}{4} F_2^2 \left[ q^2 + \omega^2 \right]; \quad \Gamma_{03} = \Gamma_{03} = -\tfrac{1}{4} F_2^2 \omega^2 . \tag{6.3}$$

Then, substituting (6.3) into (6.2) yields:

$$\slashed{q} = \left( \gamma^0 (\Gamma_{00} + \Gamma_{03}) + \gamma^3 (\Gamma_{33} + \Gamma_{30}) \right) \omega = \begin{pmatrix} \left( F_1^2 - \tfrac{1}{4} F_2^2 q^2 \right) & -\sigma^3 \left( F_1^2 - \tfrac{1}{4} F_2^2 q^2 \right) \\ \sigma^3 \left( F_1^2 - \tfrac{1}{4} F_2^2 q^2 \right) & -\left( F_1^2 - \tfrac{1}{4} F_2^2 q^2 \right) \end{pmatrix} \omega . \tag{6.4}$$

This includes a "recursive" definition of $\slashed{q}$ in terms of itself which can, it turns out, be isolated using the quadratic equation. However, the term of interest in (6.1) is $q^2$. Using (6.4) with $\sigma^3 \sigma^3 = 1$, this is found to be:

$$\slashed{q}^2 = \begin{pmatrix} \left( F_1^2 - \tfrac{1}{4} F_2^2 q^2 \right) & -\sigma^3 \left( F_1^2 - \tfrac{1}{4} F_2^2 q^2 \right) \\ \sigma^3 \left( F_1^2 - \tfrac{1}{4} F_2^2 q^2 \right) & -\left( F_1^2 - \tfrac{1}{4} F_2^2 q^2 \right) \end{pmatrix} \begin{pmatrix} \left( F_1^2 - \tfrac{1}{4} F_2^2 q^2 \right) & -\sigma^3 \left( F_1^2 - \tfrac{1}{4} F_2^2 q^2 \right) \\ \sigma^3 \left( F_1^2 - \tfrac{1}{4} F_2^2 q^2 \right) & -\left( F_1^2 - \tfrac{1}{4} F_2^2 q^2 \right) \end{pmatrix} \omega^2 = 0 . \tag{6.5}$$

This signifies that the photon is massless. Because $q^2 = 0$ (6.1) simplifies to:

$$\Gamma_{\mu\nu} = F_1^2 \eta_{\mu\nu} + \tfrac{1}{4} F_2^2 \eta_{\sigma\mu} \eta_{\nu\tau} q^\sigma q^\tau . \tag{6.6}$$

Although derived for a photon traveling along the z-axis, (6.6) is perfectly general because of its Poincaré and general coordinate invariance. This is the result we will need in section 7, as $\Gamma_{\mu\nu}$ will become our covariant metric tensor.

While we are here, however, with $q^2 = 0$ removed, we can express all components of $\Gamma_{\mu\nu}$ easily and explicitly. For a z-propagating photon, these are:

$$\Gamma_{\mu\nu} = \begin{pmatrix} F_1^2 + \frac{1}{4}F_2^2\omega^2 & 0 & 0 & -\frac{1}{4}F_2^2\omega^2 \\ 0 & -F_1^2 & 0 & 0 \\ 0 & 0 & -F_1^2 & 0 \\ -\frac{1}{4}F_2^2\omega^2 & 0 & 0 & -F_1^2 + \frac{1}{4}F_2^2\omega^2 \end{pmatrix}. \tag{6.7}$$

If we now form a tetrad from the photon polarization vectors $\varepsilon_\nu^{(\lambda)}$, $\lambda = \pm 1$, a timelike vector $\eta_\mu = (1,0,0,0)$, and a spacelike vector $\bar{q}_\mu$ as in [5], section 7.1 (Ryder denotes the latter as $\bar{k}_\mu$), and if we take $\Gamma_{\mu\nu}$ in (6.6) to be a metric tensor $g_{\mu\nu} = F_1^2\eta_{\mu\nu} + \frac{1}{4}F_2^2\eta_{\sigma\mu}\eta_{\nu\tau}q^\sigma q^\tau$, then the spin sum for the physical photon, equation (7.17) in [5], may be written:

$$\sum_{\lambda=\pm 1}\varepsilon_\mu^{(\lambda)*}\varepsilon_\nu^{(\lambda)} = -g_{\mu\nu} + \eta_\mu\eta_\nu - \bar{q}_\mu\bar{q}_\nu = -F_1^2\eta_{\mu\nu} + \eta_\mu\eta_\nu - \frac{1}{4}F_2^2\eta_{\sigma\mu}\eta_{\nu\tau}q^\sigma q^\tau - \bar{q}_\mu\bar{q}_\nu. \tag{6.8}$$

This then, in the usual manner, leads to a photon propagator:

$$\Delta_{\mu\nu} = \frac{i\sum_{\lambda=\pm 1}\varepsilon_\mu^{(\lambda)*}\varepsilon_\nu^{(\lambda)}}{q^\sigma q_\sigma + i\varepsilon} = i\frac{-F_1^2\eta_{\mu\nu} + \eta_\mu\eta_\nu - \frac{1}{4}F_2^2\eta_{\sigma\mu}\eta_{\nu\tau}q^\sigma q^\tau - \bar{q}_\mu\bar{q}_\nu}{q^\sigma q_\sigma + i\varepsilon}. \tag{6.9}$$

A similar calculation to all of the above can and should be carried out for an on-shell *massive* vector boson with momentum $P^\sigma \equiv (p'-p)^\sigma$ and mass $M^2 = P^\sigma P_\sigma$, with the caution that there is no known massive vector boson for QED, so that a complete treatment describing real physics, must necessarily go beyond QED into non-Abelian interactions such as the weak interaction. We defer this for a future paper, and leave it as well to the interested reader.

## 7. Calculation of the Magnetic Anomaly, Using the Anticommutators as a Metric Tensor

Now, we perform a calculation similar to that of section 5, but for Dirac's equation written as $(\Gamma^\mu\Gamma_{\mu\nu}p^\nu - m)u(p) = e\Gamma^\mu\Gamma_{\mu\nu}A^\nu u(p)$, where $\Gamma^\mu$ and $\Gamma_{\mu\nu}$ appear throughout in place of $\gamma^\mu$ and $\eta_{\mu\nu}$. That is, we use $\mathscr{L} = i\bar{\psi}\Gamma^\mu\Gamma_{\mu\nu}\partial^\nu\psi - e\bar{\psi}\Gamma^\mu\Gamma_{\mu\nu}A^\nu\psi - m\bar{\psi}\psi - \frac{1}{4}F^{\mu\nu}F_{\mu\nu} - A^\mu{}_{;\mu}A^\nu{}_{;\nu}$ in contrast to the $\mathscr{L} = i\bar{\psi}\gamma^\mu\eta_{\mu\nu}\partial^\nu\psi - e\bar{\psi}\Gamma^\mu\eta_{\mu\nu}A^\nu\psi - m\bar{\psi}\psi - \frac{1}{4}F^{\mu\nu}F_{\mu\nu} - A^\mu{}_{;\mu}A^\nu{}_{;\nu}$ employed in section 5. Our goal is to prove that we can obtain $\mathscr{L}(m,e,p^\mu) = \mathscr{L}'(m',e',p^{\mu\prime})$ for $A^\mu = 0$, $q^\mu = 0$, with a suitable reparameterization of $m, e, p^\mu$. We prove this equivalence, by deriving the precise mass and charge reparameterization required to render equivalent, the $A^\mu = 0$, $q^\mu = 0$, hence —p—o—p— Hamiltonians $H$ associated with the Lagrangian density $\mathcal{L}\ (m,e,p^\mu)$, and $H'$ associated with $\mathcal{L}'\ (m',e',p^{\mu'})$. In this way, it becomes possible to employ the gravitational theory in momentum space, as was developed in sections 3 and 4, with $g_{\mu\nu} = \Gamma_{\mu\nu}$, and then to simply reparameterize the $\mathcal{L}'\ (m',e',p^{\mu'})$ into $\mathcal{L}\ (m,e,p^\mu)$ to obtain the observed particle physics. To simplify notation we omit the primes for now, and will reintroduce them at the appropriate time (in (7.8), below).

We start with Dirac's equation written as $(\Gamma^\mu \Gamma_{\mu\nu}(p^\nu - eA^\nu) - m)u(p) = 0$ for a charged fermion interacting with a photon. We employ $\Gamma^\mu$ from (3.3) and the newly-derived $\Gamma_{\mu\nu} = g_{\mu\nu}$ for a photon from (6.6), to write this Dirac equation as:

$$\{[(F_1 + F_2 m)\gamma^\mu - F_2(p^\mu + \tfrac{1}{2}q^\mu)][F_1^2 \eta_{\mu\nu} + \tfrac{1}{4}F_2^2 \eta_{\sigma\mu}\eta_{\nu\tau}q^\sigma q^\tau][p^\nu - eA^\nu] - m\}u = 0. \qquad (7.1)$$

Expanding this out results in the paired equations:

$$\begin{cases} [F_1^2(F_1 + F_2 m)(E - e\phi) - F_1^2 F_2(p^\mu + \tfrac{1}{2}q^\mu)\eta_{\mu\nu}(p^\nu - eA^\nu) - m]u - F_1^2(F_1 + F_2 m)\sigma \cdot (\mathbf{p} - e\mathbf{A})v = 0 \\ F_1^2(F_1 + F_2 m)\sigma \cdot (\mathbf{p} - e\mathbf{A})u - [F_1^2(F_1 + F_2 m)(E - e\phi) + F_1^2 F_2(p^\mu + \tfrac{1}{2}q^\mu)\eta_{\mu\nu}(p^\nu - eA^\nu) + m]v = 0 \end{cases} \qquad (7.2)$$

to be contrasted with (5.7). Then, combining equations for the particle spinor $u$, and segregating the $E$ and $m$ terms, yields (contrast (5.8), and note that we thus far refrain from identifying the Hamiltonian):

$$\begin{aligned} &F_1^2(F_1 + F_2 m)E - m \\ &= F_1^2(F_1 + F_2 m)e\phi + F_1^2 F_2(p^\mu + \tfrac{1}{2}q^\mu)\eta_{\mu\nu}(p^\nu - eA^\nu) \\ &+ \frac{F_1^2(F_1 + F_2 m)^2 \sigma \cdot (\mathbf{p} - e\mathbf{A})\sigma \cdot (\mathbf{p} - e\mathbf{A})}{(F_1 + F_2 m)E + m - (F_1 + F_2 m)e\phi + F_1^2 F_2(p^\mu + \tfrac{1}{2}q^\mu)\eta_{\mu\nu}(p^\nu - eA^\nu)} \end{aligned} \qquad (7.3)$$

Next, we employ the (5.9) analog:

$$\sigma \cdot (\mathbf{p} - e\mathbf{A})\sigma \cdot (\mathbf{p} - e\mathbf{A}) = (\mathbf{p} - e\mathbf{A})^2 - ie\sigma \cdot (\mathbf{p} \times \mathbf{A} + \mathbf{A} \times \mathbf{p}) = (\mathbf{p} - e\mathbf{A})^2 - 2e\hbar \frac{\sigma}{2} \cdot \mathbf{B} \qquad (7.4)$$

to rewrite (7.3) as: (contrast (5.10))

$$F_1^2(F_1 + F_2 m)E - m$$
$$= F_1^2(F_1 + F_2 m)e\phi - F_1^2 F_2 (p^\mu + \tfrac{1}{2} q^\mu)\eta_{\mu\nu}(eA^\nu - p^\nu)$$
$$+ \frac{F_1^2(F_1 + F_2 m)^2 (\mathbf{p} - e\mathbf{A})^2}{(F_1 + F_2 m)E + m - (F_1 + F_2 m)e\phi - F_1^2 F_2 (p^\mu + \tfrac{1}{2} q^\mu)\eta_{\mu\nu}(eA^\nu - p^\nu)}$$
$$- \frac{4m F_1^2 (F_1 + F_2 m)^2}{(F_1 + F_2 m)E + m - (F_1 + F_2 m)e\phi - F_1^2 F_2 (p^\mu + \tfrac{1}{2} q^\mu)\eta_{\mu\nu}(eA^\nu - p^\nu)} \frac{e\hbar}{2m} \frac{\sigma}{2} \cdot \mathbf{B} \quad . \tag{7.5}$$

From this, we may extract the gyromagnetic ratio: (contrast (5.11))

$$\frac{g}{2} = \frac{2m F_1^2 (F_1 + F_2 m)^2}{(F_1 + F_2 m)E + m - (F_1 + F_2 m)e\phi + F_1^2 F_2 (p^\mu + \tfrac{1}{2} q^\mu)\eta_{\mu\nu}(p^\nu - eA^\nu)}. \tag{7.6}$$

Now, consider when $A^\mu = 0$, $q^\mu = 0$. From Weinberg's form factor (11.3.31) in [4], reproduced as (2.13), $q^\mu = 0$ also means $F_1(q^2 = 0) = 1$. Thus, (7.5) becomes, for $(p,q,p') = (p,0,p)$:

$$(1 + F_2 m)E - m = F_2 p^\mu \eta_{\mu\nu} p^\nu - \frac{4m(1 + F_2 m)^2}{(1 + F_2 m)E + m + F_2 p^\mu \eta_{\mu\nu} p^\nu} \frac{e\hbar}{2m} \frac{\sigma}{2} \cdot \mathbf{B}, \tag{7.7}$$

Now, we saw from (4.16), that $g_{\mu\nu}(p,0,p) = \Gamma_{\mu\nu}(p,0,p) = \eta_{\mu\nu}$, since $R^\alpha{}_{\beta\mu\nu}(p,0,p) = 0$, (4.15). So, we set $\eta_{\mu\nu} = g_{\mu\nu}$ in $p^\mu \eta_{\mu\nu} p^\nu$, thus $F_2 p^\mu \eta_{\mu\nu} p^\nu = F_2 p^\mu g_{\mu\nu} p^\nu = F_2 m^2$. We then consolidate, and in view of this extra mass term $F_2 p^\mu \eta_{\mu\nu} p^\nu = F_2 m^2$, subtract $F_2 m^2$ from each side. Now, we can identify the classical Hamiltonian as:

$$H'(p',0,p') = (1 + F_2' m')(E' - m') = -\frac{4m'(1 + F_2' m')}{E' + m'} \frac{e'\hbar}{2m'} \frac{\sigma}{2} \cdot \mathbf{B} = -g' \frac{e'\hbar}{2m'} \frac{\sigma}{2} \cdot \mathbf{B}. \tag{7.8}$$

Above, we have introduced the "prime" notation for all of the physical quantities appearing in the above, including the form factor $F_2'$ because this is a function of both $\alpha = e^2/4\pi$ and $m$, see, e.g., (2.5). Note, however, that $\mathbf{B}$ does not receive a prime denotation, because it represents the experimentally-applied magnetic field interacting with the magnetic moment. Above, distinguish $p'$ from that in $q^\mu \equiv (p'-p)^\mu$ by context.

Now, we *require* that $H'$ above be equivalent to $H$ appearing in (5.12), i.e., we *demand*:

$$H'(p',0,p') = (1 + F_2' m')(E' - m') = -\frac{4m'(1 + F_2' m')}{E' + m'} \frac{e'\hbar}{2m'} \frac{\sigma}{2} \cdot \mathbf{B} = -g' \frac{e'\hbar}{2m'} \frac{\sigma}{2} \cdot \mathbf{B}$$
$$\equiv H(p,0,p) = (\quad 1 \quad)(E - m) = -\frac{4m(1 + F_2 m)}{E + m} \frac{e\hbar}{2m} \frac{\sigma}{2} \cdot \mathbf{B} = -g \frac{e\hbar}{2m} \frac{\sigma}{2} \cdot \mathbf{B} \tag{7.9}$$

where we have also set $q^\mu = 0$ hence $F_1(q^2 = 0) = 1$ for the $H$ derived in (5.12).

Comparing, we see that the gyromagnetic terms are identical in form, but the Hamiltonian terms do not appear to be, because in (7.9) there is an additional factor of $1+F_2'm'$ multiplying $E'-m'$. In fact, this is to be expected: when all is distilled, this is the upshot of changing the kinetic term in the Lagrangian from $i\bar{\psi}\gamma^\mu \eta_{\mu\nu}\partial^\nu \psi$ to $i\bar{\psi}\Gamma^\mu \Gamma_{\mu\nu}\partial^\nu \psi$. The correction $1+F_2'm'$ now applies to $H = E - m$ as well as the gyromagnetic term. Now, we need relate $E'$, $m'$ and $e'$ to $E$, $m$ and $e$ on an individual basis, to render $H = H'$.

Referring back to (5.1) through (5.5), especially (5.2), we see that renormalization is effectively a reparameterization of various physical objects appearing in the Lagrangian where the objects being reparameterized happen to be divergent, and the coefficients (5.3) used to reparameterize are designed to be infinite as $\varepsilon \to 0$ so as to counteract these divergences. Here, we follow a similar course, but reparameterize as between *finite* objects in the Lagrangian.

First, we may render the $E - m$ terms in (7.9) equivalent if we reparameterize the mass $m$ and the energy component $E$ with:

$$m = (1+F_2'm')m'; \quad E = (1+F_2'm')E'. \tag{7.10}$$

Substituting (7.10) back into (7.9) then yields:

$$\begin{aligned} H' &= E - m = -\frac{4m(1+F_2'm')^2}{E+m}\frac{e'\hbar}{2m}\frac{\sigma}{2}\cdot\mathbf{B} = -g'\frac{e'\hbar}{2m}(1+F_2'm')\frac{\sigma}{2}\cdot\mathbf{B} \\ &\equiv H = E - m = -\frac{4m(1+F_2 m)}{E+m}\frac{e\hbar}{2m}\frac{\sigma}{2}\cdot\mathbf{B} = -g\frac{e\hbar}{2m}\frac{\sigma}{2}\cdot\mathbf{B} \end{aligned} \tag{7.11}$$

Now, the $E-m$ terms are identical, but the gyromagnetic terms are no longer so because of the substitution $m' \to m/(1+F_2'm')$ in the $\frac{e'\hbar}{2m'}$ term. To absorb this, we need to equate the middle terms using a similar reparameterization of $e$. This is achieved by requiring:

$$(1+F_2 m)e = (1+F_2'm')^2 e'. \tag{7.12}$$

Substituting this into (7.11) then yields:

$$\begin{aligned} H' &= E - m = -\frac{4m(1+F_2 m)}{E+m}\frac{e\hbar}{2m}\frac{\sigma}{2}\cdot\mathbf{B} = -g'\frac{(1+F_2 m)}{(1+F_2'm')}\frac{e\hbar}{2m}\frac{\sigma}{2}\cdot\mathbf{B} \\ &\equiv H = E - m = -\frac{4m(1+F_2 m)}{E+m}\frac{e\hbar}{2m}\frac{\sigma}{2}\cdot\mathbf{B} = -g\frac{e\hbar}{2m}\frac{\sigma}{2}\cdot\mathbf{B} \end{aligned} \tag{7.13}$$

All terms except for the final term are now equivalent. For this final step, we reparameterize the magnetic moment itself, by setting:

$$\frac{g}{(1+F_2 m)} = \frac{g'}{(1+F_2'm')} = \frac{4m}{E+m} \xrightarrow{E=m} 2, \tag{7.14}$$

so that (7.13) becomes:

$$H' = E - m = -\frac{4m(1+F_2 m)}{E+m}\frac{e\hbar}{2m}\frac{\sigma}{2}\cdot\mathbf{B} = -g\frac{e\hbar}{2m}\frac{\sigma}{2}\cdot\mathbf{B}$$
$$\equiv H = E - m = -\frac{4m(1+F_2 m)}{E+m}\frac{e\hbar}{2m}\frac{\sigma}{2}\cdot\mathbf{B} = -g\frac{e\hbar}{2m}\frac{\sigma}{2}\cdot\mathbf{B}$$
(7.15)

Thus, by reparameterizing according to (7.10), (7.12) and (7.14), the Hamiltonians $H'(p',0,p',m',e') \equiv H(p,0,p,m,e)$ are completely equivalent, describing identical observable physics, and in particular, identical, observable magnetic anomalies. These can also be combined, and expressed in terms of the gyromagnetic ratio, also shown in the $E = m$ limit, by:

$$m = \frac{E+m}{4m}g'm' \xrightarrow{E=m} m = \frac{g'}{2}m'; \quad \left(\frac{E+m}{4m}g\right)e = \left(\frac{E+m}{4m}g'\right)^2 e' \xrightarrow{E=m} \frac{g}{2}e = \left(\frac{g'}{2}\right)^2 e'.$$
(7.16)

To first order in $\alpha$, with form factors (2.5), these reparameterizations, for $E = m$, lead to:

$$m = \left(1+\frac{\alpha'}{2\pi}\right)m'; \quad \left(1+\frac{\alpha}{2\pi}\right)e = \left(1+\frac{\alpha'}{2\pi}\right)^2 e'; \quad g = 2\left(1+\frac{\alpha}{2\pi}\right),$$
(7.17)

which includes the Schwinger magnetic anomaly, as is required. And so, we have met the objective of the proof of sections 5 through 7. Because $p^0 = E = (1+F_2'm')E'$ is the time component of the vector $p^\mu$, we also may generalize (7.10) to:

$$E = (1+F_2'm')E'; \quad \mathbf{p} = (1+F_2'm')\mathbf{p}'; \quad p^\mu = (1+F_2'm')p^{\mu'}.$$
(7.18)

What we have demonstrated is the following: In the $A^\mu = 0$, $q^\mu = 0$ limit, Lagrangian density $\mathcal{L}' = i\overline{\psi}\Gamma^\mu\Gamma_{\mu\nu}\partial^\nu\psi - e'\overline{\psi}\Gamma^\mu\Gamma_{\mu\nu}A^\nu\psi - m'\overline{\psi}\psi - \tfrac{1}{4}F^{\mu\nu}F_{\mu\nu} - A^\mu{}_{;\mu}A^\nu{}_{;\nu}$ can indeed be made to describe the identical physics as $\mathcal{L} = i\overline{\psi}\gamma^\mu\eta_{\mu\nu}\partial^\nu\psi - e\overline{\psi}\Gamma^\mu\eta_{\mu\nu}A^\nu\psi - m\overline{\psi}\psi - \tfrac{1}{4}F^{\mu\nu}F_{\mu\nu} - A^\mu{}_{;\mu}A^\nu{}_{;\nu}$, *if and only if* the "primed" quantities in $\mathcal{L}'$ are related to observed quantities in $\mathcal{L}$ according to (7.10), (7.12) and (7.14). With these relationships, we do indeed obtain $H = H'$, using the Hamiltonians of section 5 and of the present section 7. Given this correspondence in the $A^\mu = 0$, $q^\mu = 0$, $\underset{p}{\longrightarrow}\!\circ\!\underset{p}{\longrightarrow}$ limit, this means that *if* the relationships (7.10), (7.12) and (7.14) hold true, then the anticommutator $\Gamma_{\mu\nu}$ derived in (2.11) *can* be used as a metric tensor $g_{\mu\nu} = \Gamma_{\mu\nu}$ for $A^\mu = 0$, $q^\mu = 0$ in $\mathcal{L}'$ to describe exactly the same physics as $\mathcal{L}$. Therefore, at least so far, $\Gamma_{\mu\nu}$ is *not falsified* from being used as a metric tensor.

To summarize most concisely, we have found that for $A^\mu = 0$, $q^\mu = 0$, i.e., $\underset{p}{\longrightarrow}\!\circ\!\underset{p}{\longrightarrow}$:

$$\mathcal{L}\left(m = (1+F_2'm')m', \quad (1+F_2 m)e = (1+F_2'm')^2 e', \quad p^\mu = (1+F_2'm')p^{\mu'}\right) = \mathcal{L}'(m',e',p^{\mu'})$$
(7.19)

or, in terms of the gyromagnetic ratio, for $E = m$:

$$\mathcal{L}\left(m=\frac{g'}{2}m',\ \frac{g}{2}e=\left(\frac{g'}{2}\right)^2 e',\ p^\mu=\frac{g'}{2}p^{\mu'}\right)=\mathcal{L}'(m',e',p^{\mu'}) \tag{7.20}$$

All of the results above, are based on considering a non-interacting fermion line $\xrightarrow{p}\circ\xrightarrow{p}$. A similar calculation to all of the above can and should be carried out for $A^\mu \neq 0$, $q^\mu \neq 0$, and therefore for the generalized vertex of Figure 2. Here, we should not expect $\mathcal{L} = \mathcal{L}'$ precisely, but rather, should expect $\mathcal{L} \cong \mathcal{L}'$ for $A^\mu \cong 0$, $q^\mu \cong 0$. Thus, any differences found between $\mathcal{L} \cong \mathcal{L}'$ can be used as the basis for experimental testing. Such a generalization to $(p,q,p')$ would need to be based on a full comparison and reparameterization as between (5.10) and (7.5), with the mass terms from $F_1^2 F_2 p^\mu \eta_{\mu\nu} p^\nu$ subtracted from each side to identify the Hamiltonian, generalizing the approach that led to (7.8). In this calculation, one needs to reparameterize more-widely, including $A^\mu$, $q^\mu$, $F_1$ and $F_2$ in addition to $m, e, p^\mu$. We leave the details of this generalization to a future paper, or, to the interested reader.

## 8. Gauge Symmetry Using the Momentum Space Metric Tensor, and the Metric Tensor as an Operator on Fermion Wavefunctions

At this point, we have derived the primary results of this paper, which are to show that, if one employs the $\Gamma_{\mu\nu} = g_{\mu\nu}$ hypothesis, then the Ward-Takahashi identities provide a natural theoretical foundation for representing QED quantum field theory in a geometrodynamic momentum space leading to an understanding of interaction as curvature, see Figure 2; and also, to show that the $\Gamma_{\mu\nu} = g_{\mu\nu}$ hypothesis, at the very least, is not falsified by the observable magnetic anomaly because this magnetic anomaly can be fully described using a metric tensor $\Gamma_{\mu\nu} = g_{\mu\nu}$ in $\mathcal{L}'$ for $A^\mu = 0$, $q^\mu = 0$, as just proven. From here, we will present several other results of interest, toward further future development.

Let us now turn back to where we left of at the end of section 4, with equation (4.16) which we now use as the operator equation $\left(g^{\mu\nu}(p,0,p) - \eta^{\mu\nu}\right)\psi = 0$. The purpose of the discussion in the current section is to provide a brief introduction to how $g^{\mu\nu}$ in momentum space can be employed to operate on Fermions. Once again, we consider the $A^\mu = 0$, $q^\mu = 0$ limit, for a non-interacting fermion line $\xrightarrow{p}\circ\xrightarrow{p}$, which also means $F_1 = 1$, see (2.13). Once again, further generalizations are possible and desirable for $A^\mu \neq 0$, $q^\mu \neq 0$, and thus for the generalized vertex of Figure 2, but we defer these to a future paper and the reader's own interest.

We begin by specifying the wavefunction for a fermion field quantum in the usual way:

$$\psi = u(p)e^{-ip^\sigma x_\sigma}. \tag{8.1}$$

With this, we use the $g^{\mu\nu}(p,0,p)$ derivable from in (3.4) to operate on $\psi$, thus writing (4.16) as the *field equation* (in this section, $(p,0,p)$ is to be understood unless stated otherwise):

$$(g^{\mu\nu} - \eta^{\mu\nu})\psi = 0 = [(2F_2 m + F_2^2 m^2)\eta^{\mu\nu} - i(1 + F_2 m)F_2(\gamma^\mu \partial^\nu + \gamma^\nu \partial^\mu) - F_2^2 \partial^\mu \partial^\nu]\psi. \qquad (8.2)$$

The final term makes use of $-F_2^2 \partial^\mu \partial^\nu \psi = iF_2^2 \partial^\mu p^\nu \psi = F_2^2 p^\mu p^\nu \psi$.

Now, as we do with Dirac's equation, we form an adjoint equation. First, we take the Hermitian conjugate of (8.2), to write:

$$\psi^\dagger (g^{\mu\nu\dagger} - \eta^{\mu\nu}) = 0$$
$$= \psi^\dagger \gamma^0 \gamma^0 (2F_2 m + F_2^2 m^2)\eta^{\mu\nu} + i(1 + F_2 m)F_2 (\partial^\nu \psi^\dagger \gamma^0 \gamma^\mu \gamma^0 + \partial^\mu \psi^\dagger \gamma^0 \gamma^\nu \gamma^0) - F_2^2 \partial^\mu \partial^\nu \psi^\dagger \gamma^0 \gamma^0 \qquad (8.3)$$

where we have employed the usual $\gamma^{\mu\dagger} = \gamma^0 \gamma^\mu \gamma^0$, as well as $1 = \gamma^0 \gamma^0$. Now, we multiply from the right by $\gamma^0$, and set $\overline{\psi} = \psi^\dagger \gamma^0$. We also observe from (3.4), that $g^{\mu\nu\dagger} = \gamma^0 g^{\mu\nu} \gamma^0$ applies generally for $(p',q,p)$ as well as for the $(p,0,p)$ limit, thus, $g^{\mu\nu\dagger} \gamma^0 = \gamma^0 g^{\mu\nu}$. With all of this, we may write (8.3) as:

$$\overline{\psi}(g^{\mu\nu} - \eta^{\mu\nu}) = 0 = [\overline{\psi}(2F_2 m + F_2^2 m^2)\eta^{\mu\nu} + i(1 + F_2 m)F_2((\partial^\nu \overline{\psi})\gamma^\mu + (\partial^\mu \overline{\psi})\gamma^\nu) - F_2^2 (\partial^\mu \partial^\nu \overline{\psi})]. \quad (8.4)$$

Then, we multiply (8.2) from the left by $\overline{\psi}$, (8.4) from the right by $\psi$, and add, to obtain:

$$2\overline{\psi}(g^{\mu\nu} - \eta^{\mu\nu})\psi = 0 = 2(2F_2 m + F_2^2 m^2)\eta^{\mu\nu} \overline{\psi}\psi$$
$$+ i(1 + F_2 m)F_2 [(\partial^\nu \overline{\psi})\gamma^\mu \psi - \overline{\psi}\gamma^\mu (\partial^\nu \psi) + (\partial^\mu \overline{\psi})\gamma^\nu \psi - \overline{\psi}\gamma^\nu (\partial^\mu \psi)]. \qquad (8.5)$$
$$- F_2^2 [\overline{\psi}(\partial^\mu \partial^\nu \psi) + (\partial^\mu \partial^\nu \overline{\psi})\psi]$$

Now, *before* going any further, let us insist on making (8.5) invariant under a *local* U(1) gauge transformation $\psi \to \psi' = e^{i\alpha(x^\mu)}\psi$. Thus, in the usual manner, we must replace $\partial^\mu$ by the gauge-covariant derivative:

$$\partial^\mu \psi \to D^\mu \psi = \partial^\mu \psi - ieA^\mu \psi; \quad \partial^\mu \overline{\psi} \to D^{\mu*} \overline{\psi} = \partial^\mu \overline{\psi} + ieA^\mu \overline{\psi}. \qquad (8.6)$$

To be clear, because we are considering $A^\mu = 0$, $q^\mu = 0$ at the moment, the gauge-covariant derivative (8.6) is in this instance reverts to the ordinary derivative, $D^\mu = \partial^\mu$. However, we will briefly carry the $ieA^\mu$ term, not because it is part of the final result here, but rather, because it illustrates how gauge symmetry is introduced into a theory which employs the momentum space metric tensor $\Gamma^{\mu\nu} = g^{\mu\nu}$.

With this, (8.5) becomes:

$$2\overline{\psi}(g^{\mu\nu} - \eta^{\mu\nu})\psi = 0 = 2(2F_2 m + F_2^2 m^2)\eta^{\mu\nu} \overline{\psi}\psi$$
$$+ i(1 + F_2 m)F_2 [(D^{\nu*}\overline{\psi})\gamma^\mu \psi - \overline{\psi}\gamma^\mu (D^\nu \psi) + (D^{\mu*}\overline{\psi})\gamma^\nu \psi - \overline{\psi}\gamma^\nu (D^\mu \psi)]. \qquad (8.7)$$
$$- F_2^2 [\overline{\psi}(D^\mu D^\nu \psi) + (D^{\mu*} D^{\nu*} \overline{\psi})\psi]$$

Now, we make explicit use of (8.6). Dividing out the factor of 2, and making use of $\partial^\mu \partial^\nu (\bar\psi \psi) = \bar\psi (\partial^\mu \partial^\nu \psi) + (\partial^\mu \partial^\nu \bar\psi) \psi + (\partial^\nu \bar\psi)(\partial^\mu \psi) + (\partial^\mu \bar\psi)(\partial^\nu \psi)$, the result is:

$$\bar\psi (g^{\mu\nu} - \eta^{\mu\nu}) \psi = 0 = (2F_2 m + F_2^2 m^2) \eta^{\mu\nu} (\bar\psi \psi)$$
$$+ \tfrac{1}{2} i(1 + F_2 m) F_2 \left[ (\partial^\nu \bar\psi) \gamma^\mu \psi - \bar\psi \gamma^\mu (\partial^\nu \psi) + (\partial^\mu \bar\psi) \gamma^\nu \psi - \bar\psi \gamma^\nu (\partial^\mu \psi) + 2ie(A^\nu \bar\psi \gamma^\mu \psi + A^\mu \bar\psi \gamma^\nu \psi) \right].  \quad (8.8)$$
$$- \tfrac{1}{2} F_2^2 \begin{bmatrix} \partial^\mu \partial^\nu (\bar\psi \psi) - (\partial^\nu \bar\psi)(\partial^\mu \psi) - (\partial^\mu \bar\psi)(\partial^\nu \psi) \\ + ieA^\mu [(\partial^\nu \bar\psi) \psi - \bar\psi (\partial^\nu \psi)] + ieA^\nu [(\partial^\mu \bar\psi) \psi - \bar\psi (\partial^\mu \psi)] - 2e^2 A^\mu A^\nu \bar\psi \psi \end{bmatrix}$$

Next, we lower one of the indexes and contract, and use $J^\mu \equiv \bar\psi \gamma^\mu \psi$, to write out the trace field equation:

$$0 = 4 [2F_2 m + F_2^2 m^2] (\bar\psi \psi)$$
$$+ \tfrac{1}{2} i(1 + F_2 m) F_2 [2(\partial_\mu \bar\psi) \gamma^\mu \psi - 2 \bar\psi \gamma^\mu (\partial_\mu \psi) + 4ieA_\mu J^\mu] \quad (8.9)$$
$$- \tfrac{1}{2} F_2^2 [\partial^\mu \partial_\mu (\bar\psi \psi) - 2(\partial_\mu \bar\psi)(\partial^\mu \psi) + 2ieA_\mu [(\partial^\mu \bar\psi) \psi - \bar\psi (\partial^\mu \psi)] - 2e^2 A^\mu A_\mu \bar\psi \psi]$$

This can further be reduced using the Dirac equation $i\gamma^\mu (\partial_\mu \psi) - m\psi = 0$ for a free Fermion and its adjoint equation $i(\partial_\mu \bar\psi) \gamma^\mu + m\bar\psi = 0$. Normally, one adds these to obtain the continuity equation $\partial_\mu (\bar\psi \gamma^\mu \psi) = 0$ for the conserved Noether current. But we can also take their difference to obtain $2m\bar\psi \psi = i\bar\psi \gamma^\mu (\partial_\mu \psi) - i(\partial_\mu \bar\psi) \gamma^\mu \psi$. Substituting this into (8.9) and consolidating, including dividing out an $F_2$ from $F_2^2$, then yields:

$$0 = 4F_2 m^2 (\bar\psi \psi) + F_2 [\tfrac{1}{2} \partial^\mu \partial_\mu (\bar\psi \psi) - (\partial_\mu \bar\psi)(\partial^\mu \psi)]$$
$$+ 2(1 + F_2 m) e A_\mu J^\mu + ieF_2 A_\mu [(\partial^\mu \bar\psi) \psi - \bar\psi (\partial^\mu \psi)] - F_2 e^2 A^\mu A_\mu \bar\psi \psi \quad (8.10)$$

We especially note the appearance of terms with the field combination $\bar\psi \psi$ which appears in fermion mass terms $m\bar\psi \psi$, as well as interaction terms $eA_\mu J^\mu$ with the fermion current density $\bar\psi \gamma^\mu \psi = J^\mu$. Most interestingly, the term $F_2 e^2 A^\mu A_\mu \bar\psi \psi$ is in the nature of a fermion mass term, yet it arises solely by appeal to local gauge symmetry. And, contained within this term, is the term $e^2 A^\mu A_\mu$, which is in the nature of a vector boson mass term (which, following electroweak spontaneous symmetry breaking, becomes zero for the photon).

As noted above, because (8.8), (8.10) above are based on the metric tensor $g^{\mu\nu}(p', q, p) = g^{\mu\nu}(p, 0, p)$, the $A^\mu$ appearing in the above is zero in any event, $A^\mu = 0$, and the graphs being described are $\xrightarrow{p} \circ \xrightarrow{p}$. Having carried the $A^\mu$ this far to illustrate how gauge symmetry may be introduced in conjunction with a momentum space metric tensor $\Gamma^{\mu\nu} = g^{\mu\nu}$, we now set $A^\mu = 0$. Thus, (8.8) and (8.10) respectively reduce to:

$$0 = (2+F_2 m)m\eta^{\mu\nu}(\overline{\psi}\psi) + \tfrac{1}{2}i(1+F_2 m)[(\partial^\nu \overline{\psi})\gamma^\mu \psi - \overline{\psi}\gamma^\mu(\partial^\nu \psi) + (\partial^\mu \overline{\psi})\gamma^\nu \psi - \overline{\psi}\gamma^\nu(\partial^\mu \psi)]$$
$$-\tfrac{1}{2}F_2[\partial^\mu \partial^\nu(\overline{\psi}\psi) - (\partial^\nu \overline{\psi})(\partial^\mu \psi) - (\partial^\mu \overline{\psi})(\partial^\nu \psi)] \quad (8.11)$$

$$0 = (\tfrac{1}{2}\partial^\mu \partial_\mu + 4m^2)(\overline{\psi}\psi) - (\partial_\mu \overline{\psi})(\partial^\mu \psi). \quad (8.12)$$

It is also of interest to obtain a Lagrangian formulation of the above. For this, we start with each of (8.2) and (8.4), respectively, in contracted form:

$$0 = +2(1+F_2 m)i\gamma^\mu(\partial_\mu \psi) - 4(2+F_2 m)m\psi + F_2(\partial^\mu \partial_\mu \psi) \quad (8.13)$$

$$0 = -2(1+F_2 m)i(\partial_\mu \overline{\psi})\gamma^\mu - 4(2+F_2 m)m\overline{\psi} + F_2(\partial^\mu \partial_\mu \overline{\psi}), \quad (8.14)$$

The Lagrangian density for both of the above is then:

$$\mathcal{L} = +2(1+F_2 m)i\overline{\psi}\gamma^\mu \partial_\mu \psi - 4(2+F_2 m)m\overline{\psi}\psi - F_2(\partial^\mu \overline{\psi})(\partial_\mu \psi), \quad (8.15)$$

which is straightforward to verify using $\left(\partial^\mu \frac{\partial}{\partial(\partial^\mu \overline{\psi})} - \frac{\partial}{\partial \overline{\psi}}\right)\mathcal{L} = 0$ and $\left(\partial_\mu \frac{\partial}{\partial(\partial_\mu \psi)} - \frac{\partial}{\partial \psi}\right)\mathcal{L} = 0$ for each of $\overline{\psi}$ and $\psi$ taken as independent field variables. Note, while terms involving $i\overline{\psi}\gamma^\mu \partial_\mu \psi$ and $m\overline{\psi}\psi$ both appear in the Lagrangian density for the Dirac equation and a term $(\partial^\mu \phi^\dagger)(\partial_\mu \phi)$ appears in the Klein-Gordon equation and is used in connection with the non-Abelian gauge group product SU(2)xU(1) to reveal vector boson masses in electroweak theory, that the Lagrangian term $(\partial^\mu \overline{\psi})(\partial_\mu \psi)$, which originates in the $p^\mu p^\nu$ term of (3.2), does *not* appear anywhere in the standard model. Let us, therefore, look more closely at this new term.

While $A^\mu = 0$ for the non-interacting $\overset{p\quad p}{\text{—}\circ\text{—}}$ approximation considered here, we can gain a sense for how this Lagrangian density looks when we require local gauge symmetry. Using (8.6), (8.15) becomes:

$$\mathcal{L} = (8m + 4F_2 m^2 + F_2 e^2 A^\mu A_\mu)\overline{\psi}\psi + F_2(\partial^\mu \overline{\psi})(\partial_\mu \psi) - 2i(1+F_2 m)\overline{\psi}\gamma^\mu(\partial_\mu \psi) - 2(1+F_2 m)eA_\mu J^\mu. \quad (8.16)$$

Using the first order $F_2 = \frac{\alpha}{2\pi}\frac{1}{m}$ from (2.5), a Fermion mass-type term above is shown to be:

$$\mathcal{L}_{\text{mass}} = \left(8 + \frac{2\alpha}{\pi} + \frac{\alpha}{2\pi}\frac{e^2 A^\mu A_\mu}{m^2}\right)m\overline{\psi}\psi, \quad (8.17)$$

and, in particular, the term $\frac{\alpha}{2\pi}\frac{e^2 A^\mu A_\mu}{m^2}m\overline{\psi}\psi$ has arisen from the new Lagrangian term $(\partial^\mu \overline{\psi})(\partial_\mu \psi)$ in (8.15), strictly by employing local gauge symmetry.

The above discussion illustrates several points for further development. First, the $\Gamma_{\mu\nu} = g_{\mu\nu}$ can be used as operators on fermion wavefunctions, just like the Dirac $\gamma^\mu$. Indeed, (8.2), written as $(g^{\mu\nu}(p,0,p) - \eta^{\mu\nu})\psi = 0$, is in the nature of a field equation which after contraction as in (8.13) is quite analogous to Dirac's equation with some perturbative coefficients and with an added term $\partial^\mu \partial_\mu \psi$. Second, once we employ the wavefunction $\psi = u(p)e^{-ip^\sigma x_\sigma}$ in (8.1) to write (8.2), we can transform over from a description of physics in momentum space, to ordinary spacetime, in the usual manner. Third, once we start to utilize terms like $\partial^\mu \psi$, gauge symmetry is easily introduced through the usual substitution $\partial^\mu \to D^\mu = \partial^\mu - ieA^\mu$, and its non-Abelian generalizations which apply, for example, to weak and strong interactions. Fourth, when we sandwich the metric tensor between two wavefunctions in the form $\bar\psi g^{\mu\nu}\psi$ as in (8.8), or form a Lagrangian as in (8.16), terms of the form $e^2 A_\sigma A^\sigma$, as well as $m\bar\psi\psi$, as well as $\bar\psi\gamma^\mu\psi = J^\mu$, as well as $eA_\mu J^\mu$, all make very natural appearances. Fifth, the term $e^2 A^\mu A^\nu (\bar\psi\psi)$ in (8.8), and its contraction $e^2 A_\sigma A^\sigma (\bar\psi\psi)$ in (8.9), (8.10), (8.16), (8.17), which is an admixture of the form for both boson and fermion masses, only comes into being because of the imposition of gauge symmetry, and *does not exist otherwise*. Because a significant challenge is to have fermion mass terms $m\bar\psi\psi$ arise in the Lagrangian, not by hand, but by appeal to local gauge symmetry in the same manner that vector boson mass terms arise in electroweak theory, and in a way that is therefore *predictive* of those masses in relation to known parameters such as the Fermi vacuum expectation value (vev) and running interaction couplings, the fact that a term involving $m\bar\psi\psi$ multiplied by a scalar $F_2 e^2 A^\mu A_\mu$ can indeed be raised into being solely by appeal to local gauge symmetry, may contribute to a better understanding of why fermions have their particular observed mass values.

Finally, as noted earlier, further generalizations are possible and desirable for $A^\mu \neq 0$, $q^\mu \neq 0$ and thus the generalized vertex of Figure 2. The most important difference is that one starts with the full equation (3.2), which contains the term $\frac{1}{4}F_2^2 (p'+p)^\mu (p'+p)^\nu$, (versus $\frac{1}{4}F_2^2 p^\mu p^\nu$ used here) and one must then employ the two separate field variables $i\partial^\mu \psi(p) = p^\mu \psi(p)$ and $-i\partial^\mu \bar\psi(p') = p^{\mu\prime} \bar\psi(p')$. The term $\frac{1}{4}F_2^2(p'+p)^\mu(p'+p)^\nu$, above, for $q^\mu = 0$, is what led to $(\partial^\mu \bar\psi)(\partial_\mu \psi)$ and then to $\frac{\alpha}{2\pi} \frac{e^2 A^\mu A_\mu}{m^2} m\bar\psi\psi$ when we considered, for illustration only, the consequences of imposing a local gauge symmetry. This full generalization based on (3.2) is deferred to a future paper and the reader's interest.

## 9. Summary and Conclusion: Geometrodynamic Measurements in Momentum Space

All of the results derived here are based on a hypothetical: what would the physics look like *if* the anticommutators $\Gamma^{\mu\nu} \equiv \frac{1}{2}\{\Gamma^\mu \Gamma^\nu + \Gamma^\nu \Gamma^\mu\} = \frac{1}{2}\{\Gamma^\mu, \Gamma^\nu\}$ derived in (2.11) and (3.2) were to be regarded as a metric tensor?

Because the anticommutators are a function of particle momentum, $\Gamma_{\mu\nu}(p,q,p')$, and not of spacetime coordinates $x^\mu$, we saw in sections 3 and 4 that the $g_{\mu\nu} = \Gamma_{\mu\nu}$ hypothesis enables us to take covariant derivatives and determine the Riemann curvature tensor $R^{\alpha}{}_{\beta\mu\nu}$ in momentum space fully analogously with how this is done in general relativity in spacetime. The Ward identity stands at the heart of this approach, because covariant differentiation in momentum space is *defined* so as to ensure the exact validity of the Ward identity. Further, based on this, we obtain a very fundamental result that equation (3.11) and its corresponding Feynman graph of Figure 2, via the Ward-Takahashi generalization, tells us that *the difference between the contracted covariant and ordinary derivatives of the momentum space metric tensor, is equal to the scattering vertex times twice the form factor $F_2$*. This, in turn, tells us that interactions produce curvature, or, equivalently, curvature is symptomatic of interactions. In a nutshell, "interaction $\propto$ curvature ." It then becomes essential to see if the $g_{\mu\nu} = \Gamma_{\mu\nu}$ hypothesis can be reconciled at least with the observed magnetic moment anomaly.

Sections 5 through 7 all serve one fundamental purpose: to show that a description of the magnetic anomaly, using $g_{\mu\nu} = \Gamma_{\mu\nu}$, can indeed be fully reconciled to the usual description based on $\mathcal{L} = i\bar{\psi}\gamma^{\mu}\eta_{\mu\nu}\partial^{\nu}\psi - e\bar{\psi}\Gamma^{\mu}\eta_{\mu\nu}A^{\nu}\psi - m\bar{\psi}\psi - \frac{1}{4}F^{\mu\nu}F_{\mu\nu} - A^{\mu}{}_{;\mu}A^{\nu}{}_{;\nu}$, wherein $\Gamma^{\mu}$ appears *only* in the $e\bar{\psi}\Gamma^{\mu}\eta_{\mu\nu}A^{\nu}\psi$ term and the metric tensor is taken to be $g_{\mu\nu} = \eta_{\mu\nu}$. The $g_{\mu\nu} = \Gamma_{\mu\nu}$ hypothesis requires that we use $\mathcal{L}' = i\bar{\psi}\Gamma^{\mu}\Gamma_{\mu\nu}\partial^{\nu}\psi - e\bar{\psi}\Gamma^{\mu}\Gamma_{\mu\nu}A^{\nu}\psi - m\bar{\psi}\psi - \frac{1}{4}F^{\mu\nu}F_{\mu\nu} - A^{\mu}{}_{;\mu}A^{\nu}{}_{;\nu}$, which *appears*, superficially, to be a different Lagrangian density. Sections 5 through 7 *prove*, however, that the observable physics described by these two Lagrangian densities can be made completely identical in the $q^{\mu} = 0$ limit, thus giving us the baseline for generalizations to $q^{\mu} \cong 0$ and even to $q^{\mu} \gg 0$. In terms of the Hamiltonians derived at great length, these sections show that $H'(p^{\mu\prime}, 0, p^{\mu\prime}; m', e') \equiv H(p^{\mu}, 0, p^{\mu}; m, e)$. In particular, sections 5 through 7 derive, in detail, exactly how the physical quantities expressed in one Lagrangian must map into those expressed in the other – what we have called "reparameterization" – in order to achieve $\mathcal{L}' = \mathcal{L}$ for $q^{\mu} = 0$.

In section 8, we examined the use of the metric tensor as an operator on Fermions, and showed how this overall approach can be rendered fully compatible with established principles of local gauge symmetry, and may be helpful in better understanding the origin of fermion masses using local gauge symmetry.

Through all of this, we have, in a certain sense, backed into general relativity, expressed via a metric tensor in the momentum space of particle physics and quantum field theory. We simply calculated the mathematical anticommutators $\Gamma^{\mu\nu} \equiv \frac{1}{2}\{\Gamma^{\mu}\Gamma^{\nu} + \Gamma^{\nu}\Gamma^{\mu}\} = \frac{1}{2}\{\Gamma^{\mu}, \Gamma^{\nu}\}$ of the convergent perturbative $\Gamma^{\mu} \equiv \gamma^{\mu} + \Lambda^{\mu}$, asked what the physics would look like *if* we employed this strictly mathematical entity $\Gamma_{\mu\nu}$ as a metric tensor $g_{\mu\nu} = \Gamma_{\mu\nu}$, and required that the Ward identity be enforced and that the low-energy approximation be identical to what we know is observed in relation to the magnetic anomaly. The rest is sheer calculation. Now, the frontal question of general relativity arises: in spacetime, we know how to take proper time and distance measurements using geometrodynamic clocks and rods. How do we take similar measurements when $g_{\mu\nu}$ is specified in momentum space?

In spacetime, the differential spacetime elements in a specified choice of coordinates are $dx^{\mu}$, and the invariant proper length or time is $(ds)^2 = g_{\mu\nu}(x^{\sigma})dx^{\mu}dx^{\nu}$, where $g_{\mu\nu}(x^{\sigma})$ is a function of the spacetime coordinates $x^\sigma$. We know what it means to go from having events occur on a spacetime manifold, to using a set of coordinates to map out these events, to experimentally measuring proper times and distances between those events using geometrodynamic clocks and rods. By analogy, in momentum space, it appears as though we should consider differential momentum elements $dp^\mu$ expressed in some choice of coordinates, a metric tensor $g_{\mu\nu}(p,q,p')$ as has been developed and explored in depth here, and, in lieu of $ds$, an invariant differential mass element $dm$ specified by:

$$(dm)^2 = dp^\mu g_{\mu\nu} dp^\nu. \tag{9.1}$$

Now, so far, (9.1) is merely an analogy, not a theory of how to take measurements in momentum space. Of course, it helps that we have a candidate metric tensor $g_{\mu\nu}(p,q,p')$ which has already been explored and developed in detail. So, let's follow this along a little further.

If the momentum space invariant is to be $(dm)^2 = dp^\mu dp_\mu$, let's consider what it means to integrate over the element $dm$, that is, to take, say, $\int_{m_0}^{\infty} \frac{1}{m} dm$. Such integral is given by:

$$\int_{m_0}^{\infty} \frac{1}{m} dm = \ln m \Big|_{m_0}^{\infty} = \ln m \Big|_{m_0}^{\mu} + \ln m^2 \Big|_{\mu}^{\infty} = \ln\left(\frac{\mu}{m_0}\right) + \ln\left(\frac{\infty}{\mu}\right) = \ln\left(\frac{\mu}{m_0}\right) + \infty. \tag{9.2}$$

This most basic integral involving $(dm)^2 = dp^\mu dp_\sigma$ yields a term $\ln\left(\frac{\mu}{m_0}\right)$ which is the fundamental "probe" term used to specify a particle physics, QFT "observation," as well as a logarithmically-divergent infinity just waiting to be "swept under the rug." Of course, it is simpler just to write (9.2) without the infinity as:

$$\int_{m_0}^{\mu} \frac{1}{m} dm = \int_{m_0}^{\mu} \frac{1}{m} \sqrt{dp^\mu \Gamma_{\mu\nu} dp^\nu} = \ln m \Big|_{m_0}^{\mu} = \ln\left(\frac{\mu}{m_0}\right). \tag{9.3}$$

This is how we take measurements based on an invariant element $dm$ in momentum space. The "probe factor" $\ln\left(\frac{\mu}{m_0}\right) = \int_{m_0}^{\mu} \frac{1}{m} dm$, specified from $(dm)^2 = dp^\mu g_{\mu\nu} dp^\nu$, integrated over a specified range of energies, becomes the "clock" or "rod" used to take measurements in momentum space.

Using (9.3) and $\alpha = e^2/4\pi$, we may then rewrite the form factor $F_1(q^2)$ in (2.13), as:

$$F_1(q^2) \cong 1 + \frac{\alpha}{3\pi}\left(\frac{q^2}{m_0^2}\right)\left[\int_{m_0}^{\mu} \frac{1}{m} dm + \frac{1}{5} + \frac{3}{8}\right] = 1 + \frac{\alpha}{3\pi}\left(\frac{q^2}{m_0^2}\right)\left[\int_{m_0}^{\mu} \frac{1}{m} \sqrt{dp^\mu \Gamma_{\mu\nu} dp^\nu} + \frac{1}{5} + \frac{3}{8}\right], \tag{9.4}$$

directly incorporating the invariant element $dm$ and the anticommutators $\Gamma_{\mu\nu}$ which we have developed here, in detail, as a metric tensor $g_{\mu\nu} = \Gamma_{\mu\nu}$. In sum, in momentum space, an invariant element $(dm)^2 = dp^\mu \Gamma_{\mu\nu} dp^\nu$ lends itself naturally to taking measurements, by virtue of the usual "probe factor" (9.3), and the form factor (9.4) which appears throughout $\Gamma_{\mu\nu}$ and $\Gamma^\mu$.

In concluding, we stop short of asserting $g_{\mu\nu} = \Gamma_{\mu\nu}$ as physical reality, but for now, retain this as a *provisional hypothesis*, which, if valid, would lead to all of the results here, as well as to perhaps other results not yet explored here. We have tried here to eliminate a most important objection, by showing that a Lagrangian density in which $\Gamma_{\mu\nu}$ is taken to be the metric tensor and in which $\Gamma^\mu$ appears at all vertexes, can be made entirely consistent with the Lagrangian density customarily employed in quantum field theory which places the perturbative correction at the fermion / boson vertex only, and leads to the exact same specification of the magnetic anomaly in the $q^\mu = 0$ limit. We have also shown how the Ward-Takahashi identities lie at the heart of the formal development, and how the $g_{\mu\nu} = \Gamma_{\mu\nu}$ hypothesis is fully compatible with gauge theory and may help elucidate the problem of why the elementary fermions have the masses they have.

If further investigations of this $g_{\mu\nu} = \Gamma_{\mu\nu}$ hypothesis turn up no fundamental contradictions, and especially if they should lead to explanations of observations not yet explained or to a more accurate description of what is observed, then the hypothesis that $g_{\mu\nu} = \Gamma_{\mu\nu}$ may end up commending itself as physical reality. This would mean, among other things, that perturbative corrections can be given a parallel, isomorphic explanation in terms of gravitational effects emanating from a metric tensor $g_{\mu\nu} = \Gamma_{\mu\nu} \equiv \tfrac{1}{2}\{\Gamma_\mu \Gamma_\nu + \Gamma_\nu \Gamma_\mu\}$, and that quantum field theory may well yet find itself in a seamless marriage with geometrodynamic gravitational theory.